\documentclass[manuscript,screen]{acmart}

\usepackage{multirow} 
\usepackage{subcaption} 
\usepackage[dvipsnames]{xcolor} 
\usepackage{ifthen} 
\usepackage{framed}

\newcommand{\eoan}[1]{\textcolor{Black}{#1}}

\AtBeginDocument{%
  }

\setcopyright{acmlicensed}
\copyrightyear{2018}
\acmYear{2018}
\acmDOI{XXXXXXX.XXXXXXX}
\acmConference[Conference acronym 'XX]{Make sure to enter the correct
  conference title from your rights confirmation email}{June 03--05,
  2018}{Woodstock, NY}
\acmISBN{978-1-4503-XXXX-X/2018/06}

\begin{document}

\title{Energy Efficiency in Microservice Architectures: A Systematic Literature Review}

\author{Eoan O'Dea}
\email{e.odea@rug.nl}
\orcid{0009-0009-0469-1092}
\affiliation{%
  \institution{University of L'Aquila}
  \city{L'Aquila}
  \country{Italy,}
  \institution{University of Groningen}
  \city{Groningen}
  \country{The Netherlands}
}

\author{Roberta Capuano}
\email{roberta.capuano@univaq.it}
\orcid{0000-0001-9903-999X}
\affiliation{%
  \institution{University of L'Aquila}
  \city{L'Aquila}
  \country{Italy}}
\email{roberta.capuano@univaq.it}

\author{Paris Avgeriou}
\email{p.avgeriou@rug.nl}
\orcid{0000-0002-7101-0754}
\affiliation{%
  \institution{University of Groningen}
  \city{Groningen}
  \country{The Netherlands}}

\author{Henry Muccini}
\email{henry.muccini@univaq.it}
\orcid{0000-0001-6365-6515}
\affiliation{%
  \institution{University of L'Aquila}
  \city{L'Aquila}
  \country{Italy}}
\email{henry.muccini@univaq.it}

\renewcommand{\shortauthors}{O'Dea et al.}

\begin{abstract}
\textbf{Context.} Microservice architectures are widely adopted for building scalable cloud-native systems, enabling independent deployment, fine-grained service composition, and operational elasticity.
Problem. Despite growing interest in sustainable software, research on energy efficiency in microservices spans operational, infrastructural, and architectural perspectives, but these are typically addressed in isolation. Existing studies focus on optimisation techniques or measurement approaches, with limited synthesis of how energy efficiency is considered, measured, and addressed at the architectural level.
\textbf{Goal.} This study synthesises research on energy-efficient microservices by examining where energy efficiency is considered, how it is measured, and which architectural solutions have been proposed.
\textbf{Method.} We conduct a systematic literature review following Kitchenham’s guidelines, screening publications from four major digital libraries through a six-stage process with backward and forward snowballing, resulting in 40 primary studies.
\textbf{Results.} Energy efficiency is predominantly addressed at runtime through monitoring, scheduling, and resource management, while design-time integration remains limited. Measurement practices are largely infrastructure-oriented and rely on model-based estimation and coarse-grained monitoring.
\textbf{Conclusion.} Energy efficiency in microservices is primarily treated as an operational optimisation problem rather than a lifecycle-spanning architectural concern, highlighting the need for earlier architectural integration and improved measurement practices.

\end{abstract}




\maketitle

\newcommand{\defrqparam}[2]{%
\expandafter\def\csname rqparamname@#1\endcsname{#2}%
}

\newcommand{\rqparam}[1]{%
\ifcsname rqparamname@#1\endcsname
\textit{\csname rqparamname@#1\endcsname\ (\MakeUppercase{#1})}%
\else
\textbf{[UNKNOWN PARAMETER: #1]}%
\fi
}

\newcommand{\rqparambf}[1]{%
\ifcsname rqparamname@#1\endcsname
\textbf{\csname rqparamname@#1\endcsname\ (\MakeUppercase{#1})}%
\else
\textbf{[UNKNOWN PARAMETER: #1]}%
\fi
}

\defrqparam{rq1.1.p1}{Energy Consideration Stage}
\defrqparam{rq1.1.p2}{Energy Intervention Stage}
\defrqparam{rq1.1.p3}{Energy Optimisation Technique}

\defrqparam{rq1.2.p1}{Component Type}
\defrqparam{rq1.2.p2}{Energy consumption source}
\defrqparam{rq1.2.p3}{Component Relationship}

\defrqparam{rq2.1.p1}{Measurement Level}
\defrqparam{rq2.1.p2}{Measurement Technique}
\defrqparam{rq2.1.p3}{Data Collection}
\defrqparam{rq2.1.p4}{Implementation Approach}

\defrqparam{rq2.2.p1}{Tool Type}
\defrqparam{rq2.2.p2}{Deployment Method}
\defrqparam{rq2.2.p3}{Analysis Capability}
\defrqparam{rq2.2.p4}{Granularity}

\defrqparam{rq2.3.p1}{Metric Type}
\defrqparam{rq2.3.p2}{Measurement Scope}
\defrqparam{rq2.3.p3}{Normalization Basis}
\defrqparam{rq2.3.p4}{Business Relevance}

\defrqparam{rq3.1.p1}{Energy Management Strategy}
\defrqparam{rq3.1.p2}{Tactic Category}
\defrqparam{rq3.1.p3}{Implementation Technologies}

\defrqparam{rq3.2.p1}{Quality Attribute}
\defrqparam{rq3.2.p2}{Trade-off Relationship}
\defrqparam{rq3.2.p3}{Trade-off Strategy}

\defrqparam{rq3.3.p1}{Challenge Type}
\defrqparam{rq3.3.p2}{Adoption Phase}
\defrqparam{rq3.3.p3}{Challenge Complexity}
\defrqparam{rq3.3.p4}{Mitigation Approach}


\section{Introduction}\label{sc:intro}

Microservice architectures have emerged over the past decade as a dominant paradigm for building large-scale, distributed software systems~\cite{pahl2016microservices}. Evolving from service-oriented architectures and influenced by DevOps and continuous delivery practices, microservices decompose applications into small, autonomous services that can be developed, deployed, and scaled independently~\cite{chen2018microservices}. This architecture style has been widely adopted in cloud-native environments due to its support for organisational scalability, rapid release cycles, and operational elasticity~\cite{alshuqayran2016systematic}. These characteristics have driven widespread adoption of microservice architectures in industry, with a survey of practitioners conducted by \textit{Gartner} reporting that 74\% of respondents indicated that their organisations already use microservices, and a further 23\% planning adoption~\cite{gartnerMicroservices}. Organisations adopt microservices either when developing new systems or when re-architecting existing monolithic applications~\cite{taibi2017processes,capuano2022systematic}. Recent studies have also reported cases where organisations revert to modular monolithic architectures due to operational complexity or maintenance overhead~\cite{ponce2019migrating,taibiback2023,taibiback2024}.

Prior research highlights scalability, performance, maintainability, and deployability as commonly cited motivations for adopting microservice architectures~\cite{di2017research}. Autoscaling policies, container orchestration platforms, and fine-grained service boundaries are frequently used to improve responsiveness under variable workloads~\cite{toffetti2015architecture,yu2020microscaler}. At the same time, architectural principles such as loose coupling and high cohesion are used to enhance evolvability and team autonomy~\cite{taibi2018architectural}. As a result, microservices are commonly framed as an architecture supporting agility and operational flexibility~\cite{chen2018microservices}. However, these same characteristics may also influence system-level resource consumption~\cite{araujo2024energy}. Increased inter-service communication, containerisation overhead, replication strategies, and runtime adaptation mechanisms introduce additional infrastructure and coordination costs~\cite{zhu2022dissecting,berry2024worth}. Unlike monolithic architectures, microservices emphasise distribution and continuous operation, which can increase compute, networking, and orchestration demands, thereby affecting overall energy consumption~\cite{beloglazov2011taxonomy,mastelic2014cloud}. Consequently, while microservices are typically evaluated against performance and scalability objectives, their energy efficiency has received comparatively limited attention~\cite{araujo2024energy,xiao2024architectural}.

\subsection{Motivation and Paper Contribution}

Sustainability has recently emerged as an important non-functional requirement in software systems~\cite{freitag2021real, gill2018taxonomy}. Regulatory initiatives, particularly within the European Union, are placing increasing emphasis on energy efficiency and environmental impact in ICT infrastructures~\cite{espr,csrd}. While these concerns have traditionally been addressed at the hardware and data-centre level~\cite{bharany2022systematic}, there is increasing recognition that architecture decisions at the software level also influence long-term energy behaviour~\cite{seo, capuano2025comparative}. 

Despite the growing attention, research on microservice architectures has largely examined energy consumption from operational or infrastructure-oriented perspectives, focusing on measurement techniques, runtime optimisation, or resource management mechanisms~\cite{araujo2024energy,xiao2024architectural, bharany2022systematic}. At the same time, the broader literature on microservices is fragmented across different perspectives, including architectural design, performance optimisation, and architectural patterns~\cite{alshuqayran2016systematic,di2017research,valdivia2020patterns}. This fragmentation makes it difficult to understand how energy efficiency is addressed within microservice architectures research and how architectural decisions influence energy-related outcomes. This raises a fundamental question: \textit{How should microservice architectures be designed, evaluated, and evolved to account for energy efficiency as a first-class concern?} To address this question, this paper presents a systematic literature review of energy-efficient microservices, with the following contributions:

\begin{itemize}
\item A systematic analysis of where energy efficiency is considered in microservice architecture research, including lifecycle stages and architectural elements involved in energy-related analysis.
\item A synthesis of how energy consumption is measured in microservices, covering measurement methods, tools, and metrics.
\item A review of architectural solutions proposed to improve energy efficiency in microservices, including design approaches, trade-offs with other quality attributes, and barriers to adoption.
\item An integrated cross-analysis of these dimensions, highlighting research gaps and identifying open challenges for the development of energy-efficient microservice architectures.
\end{itemize}

\subsection{Paper Structure}

The remainder of this paper is organised as follows. Section~\ref{sc:sota} presents the background through related work on energy efficiency and microservice architectures, and positions this study with respect to prior secondary studies. Section~\ref{sc:method} describes the six-stage review process adopted in this study. Sections~\ref{sc:rq1}--\ref{sc:rq3} present the results of the review by addressing the three research questions. 
Section~\ref{sc:discussion} synthesises the findings across the research questions. Section~\ref{sc:threats} discusses threats to validity, and Section~\ref{sc:conclusion} concludes the paper.

\section{Background and Related Work}\label{sc:sota}

This section situates the review within existing research. Microservice architectures are typically analysed through architectural quality attributes and trade-offs, with performance, scalability, maintainability, availability, and reliability as the most frequently examined concerns. Architectural patterns and tactics are evaluated based on how they balance these competing qualities~\cite{alshuqayran2016systematic,di2017research,li2021understanding,de2020method,el2019guiding,lytra2020quality}. However, energy efficiency and broader sustainability concerns are rarely treated as primary architectural drivers within this body of work.

Microservice architectures also exhibit strong operational dynamism, with autoscaling, self-management, and continuous deployment embedding runtime adaptation mechanisms into the system~\cite{toffetti2015architecture,yu2020microscaler,chen2018microservices,o2017continuous}. As a result, architectural decision-making extends beyond design-time structure into deployment and runtime control. This provides the context in which energy efficiency can be examined as both an architectural and operational concern. The following sections examine how energy efficiency has been addressed in distributed systems and position this review relative to existing secondary studies.

\subsection{Energy Efficiency in Distributed Systems}

Energy efficiency in microservices builds on earlier research in distributed and cloud computing, which examined processor power states, dynamic voltage and frequency scaling (DVFS), and resource management in clusters and data centres~\cite{orgerie2014survey,beloglazov2011taxonomy,mastelic2014cloud}, framing energy efficiency primarily as an infrastructure-level optimisation problem. In microservices, this optimisation-oriented view persists. Energy reduction is commonly addressed through autoscaling, brownout control, and orchestration mechanisms in Kubernetes, edge, and serverless platforms~\cite{xu2019brownoutcon,li2025energy,wang2025energy,aslanpour2022energy}, reflecting earlier work on virtual machine placement and multi-objective resource provisioning under performance and QoS constraints~\cite{beloglazov2010energy,malekloo2018energy}. 
Even at finer granularity, energy is predominantly treated as a runtime control objective. 

Measurement practices follow a similar trajectory. Techniques such as RAPL-based instrumentation and regression power models~\cite{hackenberg2013power,mobius2013power} have been adapted to container and service-level contexts~\cite{araujo2024energy,amaral2023kepler,legler2025service}. Although recent studies analyse microservice granularity and architectural patterns in relation to energy consumption~\cite{xiao2025effectiveness,zhao2025does,berry2024worth}, \eoan{the extent to which energy efficiency is integrated into architectural design remains unclear. Energy efficiency in microservices is therefore often approached through optimisation and instrumentation techniques, motivating further investigation into its role as an architectural concern.}

    
\subsection{Existing Literature Reviews and Positioning of This Review}

Several systematic reviews examine microservice architectures from a structural and quality-oriented perspective. Studies such as Alshuqayran et al.~\cite{alshuqayran2016systematic}, Di Francesco et al.~\cite{di2019architecting}, and Valdivia et al.~\cite{valdivia2020patterns} analyse architectural decomposition, design patterns, and quality attributes including performance, scalability, maintainability, and reliability. While these reviews provide detailed insights into architectural decision-making and quality trade-offs, energy efficiency and sustainability are not treated as primary architectural concerns. 

More recently, a small number of secondary studies have focused explicitly on energy in microservices.  Araújo et al.~\cite{araujo2024energy} synthesise empirical research on energy consumption in microservices, while Xiao~\cite{xiao2024architectural} reviews architectural tactics for improving environmental sustainability. These reviews suggest growing interest in energy-efficient microservices; however, their analyses remain largely technique- and metric-oriented, with limited integration of structural design decisions, runtime mechanisms, and quality trade-offs into a unified architectural perspective. 

In parallel, broader surveys of energy efficiency in distributed and cloud computing concentrate on infrastructure-level optimisation strategies such as virtualisation, consolidation, and resource management~\cite{gill2018taxonomy,bharany2022systematic,lin2024systematic}. In these reviews, energy is primarily framed as a scheduling and resource-management objective at the data-centre level rather than as an application-level architectural design issue. \eoan{To enable comparison across studies, we organise prior work along three key dimensions: (i) how systems are structured and evolve across the lifecycle, (ii) how energy consumption is measured and evaluated, and (iii) which architectural solutions and trade-offs are considered.} Table~\ref{tab:slr_comparison} summarises the scope and analytical focus of the aforementioned secondary studies in relation to this review. The comparison highlights that prior work either \eoan{treats energy as one concern among many, often focusing on specific techniques or metrics rather than positioning energy efficiency as a primary architectural driver,} or addresses sustainability primarily at the infrastructure level. A systematic integration of architectural structure, energy measurement practices, lifecycle positioning, and quality trade-offs within microservices remains underdeveloped.

\begin{table*}[b]
\centering
\small
\renewcommand{\arraystretch}{1.2}
\begin{tabular}{p{6.5cm} p{.5cm} p{.7cm} p{1.6cm} p{1.5cm} p{1.5cm}}
\hline
\textbf{Study} & \textbf{MS} & \textbf{Energy} & \textbf{Lifecycle / Elements} & \textbf{Measurement / Metrics} & \textbf{Solutions / Trade-offs} \\
\hline

A Systematic Mapping Study in Microservice Architecture~\cite{alshuqayran2016systematic}
& Yes & No & Yes & No & Yes \\

Architecting with Microservices: A Systematic Mapping Study~\cite{di2019architecting}
& Yes & No & Yes & No & Yes \\

Patterns Related to Microservice Architecture: A Multivocal Review~\cite{valdivia2020patterns}
& Yes & No & Yes & No & Limited \\

Energy Consumption in Microservices Architectures: A Systematic Literature Review~\cite{araujo2024energy}
& Yes & Yes & Limited & Yes & Limited \\

Architectural Tactics to Improve the Environmental Sustainability of Microservices~\cite{xiao2024architectural}
& Yes & Yes & Partial & Limited & Limited \\

A Taxonomy and Future Directions for Sustainable Cloud Computing~\cite{gill2018taxonomy}
& No & Yes & No & Yes & Limited \\

Energy-Efficient Techniques in Sustainable Cloud Computing~\cite{bharany2022systematic}
& No & Yes & No & Yes & Limited \\

Green-Aware Management Techniques for Sustainable Data Centers~\cite{lin2024systematic}
& No & Yes & No & Yes & Limited \\

\textbf{This Review}
& \textbf{Yes} & \textbf{Yes} & \textbf{Yes} & \textbf{Yes} & \textbf{Yes} \\

\hline
\end{tabular}
\caption{Comparison of existing secondary studies and positioning of this review.}
\label{tab:slr_comparison}
\end{table*}

\section{Study Design}\label{sc:method}

This study follows the systematic literature review (SLR) guidelines proposed by Kitchenham et al.~\cite{kitchenham2007guidelines}. The planning and conducting phases are described in this section, while the reporting phase is reflected in the results and synthesis sections (Sections \ref{sc:rq1}-\ref{sc:discussion}).

\subsection{Planning the Review}
\begin{table}[b]
\centering
\small
\begin{tabular}{c l}
\hline
\textbf{ID} & \textbf{Research Question} \\
\hline
\addlinespace[0.2em]
\textbf{RQ1} & \textbf{Where is energy efficiency considered in microservice architecture research?} \\
\addlinespace[0.1em]
\hline 
\addlinespace[0.2em]
RQ1.1 & \hspace{0.3cm}At which stage of the software lifecycle is energy efficiency considered in microservices? \\
\addlinespace[0.2em] RQ1.2 & \hspace{0.3cm}Which microservice architectural elements are primarily considered when measuring energy consumption? \\
\addlinespace[0.2em]\hline
\addlinespace[0.2em]\textbf{RQ2} & \textbf{How is energy consumption measured in microservices?} \\ 
\addlinespace[0.1em]\hline 
\addlinespace[0.2em] RQ2.1 & \hspace{0.3cm}What methods are used to measure energy consumption in microservices? \\
\addlinespace[0.2em] RQ2.2 & \hspace{0.3cm}How are tools used for energy consumption analysis in microservices? \\
\addlinespace[0.2em] RQ2.3 & \hspace{0.3cm}What energy efficiency metrics are commonly used to evaluate microservices? \\
\addlinespace[0.2em]\hline
\addlinespace[0.2em]
\textbf{RQ3} & \textbf{What architectural solutions exist for energy-efficient microservices?} \\ \addlinespace[0.1em]\hline 
\addlinespace[0.2em] RQ3.1 & \hspace{0.3cm}What approaches are used to improve energy efficiency in microservice architectures? \\
\addlinespace[0.2em] RQ3.2 & \hspace{0.3cm}What trade-offs exist between energy efficiency and other quality attributes in microservice architectures? \\
\addlinespace[0.2em] RQ3.3 & \hspace{0.3cm}What barriers and challenges affect the adoption of energy-efficient architectural solutions in microservices? \\
\addlinespace[0.2em]\hline
\end{tabular}
\caption{Research Questions}
\label{tab:rqs}
\end{table}

The planning phase starts with the definition of the research questions, which frame the scope and objectives of the review. The proposed research questions focus on three main aspects of energy efficiency in microservices: architectural considerations, the measurement of energy consumption, and architectural solutions for improving energy efficiency. The full list of research questions is reported in Table~\ref{tab:rqs}. The search terms and their synonyms were derived from the research questions to ensure coverage of the investigated aspects. While explicit architectural terms (for RQ1) were not included in the search query, the combination of microservice-related and energy-related keywords targeted studies discussing energy considerations in microservices. Architectural aspects relevant to \textit{RQ1} were subsequently identified during the screening and data extraction phases. Based on the selected terms and their synonyms, the following search string was constructed:

\begin{center}
{\small
\textit{("microservice" OR "micro-service" OR "microservices" OR "micro-services")
AND ("energy consumption" OR "power consumption" OR "energy efficiency")
AND ("measurement" OR "measurements" OR "profiling" OR "monitoring")
AND ("tradeoff" OR "trade-off" OR "quality" OR "qualities" OR "pattern" OR "patterns" OR "tactic" OR "tactics")}}
\end{center}

We conducted the search across four major digital libraries: \textit{ACM Digital Library}, \textit{IEEE Xplore}, \textit{Scopus}, and \textit{SpringerLink}. Minor adaptations of the search string were applied to accommodate library-specific syntax and query constraints. The exact queries used for each library are available in the replication package~\cite{odea2026slrreplication}.

\subsection{Conducting the Review}

To ensure comparability, the search was restricted to English-language journal and conference papers published from 2015 onward, reflecting the period in which microservice architectures became widely adopted~\cite{di2019architecting}. Full-text queries were used because some digital libraries do not support restricting searches to titles and abstracts.

\begin{table*}[b]
\centering
\small
\renewcommand{\arraystretch}{1.2}
\begin{tabular}{p{.7cm} p{10cm} p{2.5cm}}
\hline
\textbf{ID} & \textbf{Criterion Description} & \textbf{Applied Stage} \\
\hline

\multicolumn{3}{l}{\textbf{Inclusion Criteria}} \\
\hline

IC1 & Paper published between 2015 and 2025 & Dataset \\
IC2 & Paper is a peer-reviewed conference or journal paper & Dataset \\
IC3 & Paper is not a duplicate across multiple sources & Dataset \\

IC4 & Paper written in English based on title and abstract & Preliminary \\

IC5 & Title/abstract indicates energy efficiency or consumption in a software engineering or architecture context & Primary \\
IC6 & Title/abstract refers to solutions for energy-efficient microservices & Primary \\

IC7 & Introduction/conclusion discusses engineering energy-efficient microservices (RQ1) & Secondary \\
IC8 & Introduction/conclusion discusses measurement of energy consumption in microservices (RQ2) & Secondary \\
IC9 & Introduction/conclusion refers to solutions for energy-efficient microservices (RQ3) & Secondary \\

\hline
\multicolumn{3}{l}{\textbf{Exclusion Criteria}} \\
\hline

EC1 & Paper is a graduate thesis or project report & Dataset \\
EC2 & Paper has not been peer-reviewed & Dataset \\

EC3 & Title or abstract does not mention (cloud application or microservices) and (energy, sustainability, or green) & Preliminary \\

EC4 & Study focuses only on hardware or infrastructure energy concerns & Primary + Secondary \\
EC5 & No mention of measurement, analysis, or evaluation of energy in microservices & Primary + Secondary \\

EC6 & Paper is a secondary study (SLR or mapping study) & Primary \\

EC7 & Full text not accessible & Secondary \\

\hline
\end{tabular}
\caption{Inclusion and exclusion criteria applied during the screening stages of the review.}
\label{tab:screening_criteria}
\end{table*}

\begin{figure}[t]
    \centering
    \includegraphics[width=1\linewidth]{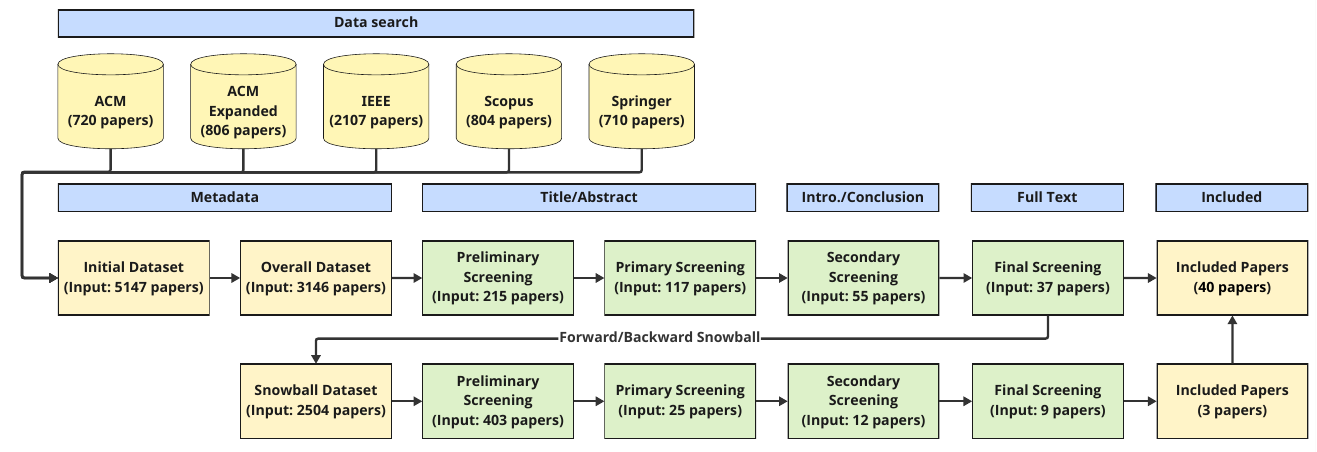}
    \caption{Overview of the systematic literature review process, including database search, multi-stage screening, and snowballing leading to the final set of included papers}
    \label{fig:research_methodology_overview}
\end{figure}

A six-stage screening process with corresponding inclusion and exclusion criteria was defined to screen the candidate studies, as illustrated in Figure~\ref{fig:research_methodology_overview}. The figure summarises the overall review workflow, including the database search, the multi-stage screening process, and the complementary backwards and forward snowballing applied after the final screening stage, which followed the guidelines proposed by Wohlin~\cite{wohlin2014guidelines} and was conducted to identify additional relevant studies that may not have been captured originally. The inclusion and exclusion criteria applied across the screening stages are summarised in Table~\ref{tab:screening_criteria}. The complete review protocol is available in the replication package~\cite{odea2026slrreplication}.

The \textbf{data search stage} involved aggregating and preprocessing the retrieved records to construct the datasets \textit{Initial Dataset} and \textit{Overall Dataset}, respectively (see Figure~\ref{fig:research_methodology_overview}). The search across \textit{ACM Digital Library}, \textit{IEEE Xplore}, \textit{Scopus}, and \textit{SpringerLink} returned 5,147 publications. Publication year (2015–2025), language, and venue filters were applied during database queries when supported by the digital libraries and otherwise during dataset preprocessing. Duplicate records across libraries were removed, resulting in a dataset of 3,146 publications used for screening, with dataset-level inclusion and exclusion criteria applied as defined in Table~\ref{tab:screening_criteria}. Screening criteria were applied progressively across stages, with broader criteria used during preprocessing and increasingly specific criteria applied in later screening phases as more information from the studies became available.

The \textbf{preliminary screening stage} applied a lightweight keyword-based relevance filter to paper titles and abstracts using the preliminary screening criteria. This step was necessary because database searches were executed across all metadata fields, as some digital libraries do not support restricting queries to titles and abstracts. Papers were retained when the title or abstract indicated a relationship between microservices or cloud applications and energy-related concerns such as energy consumption, sustainability, or green computing. The screening was conducted using the ASReview active-learning tool~\cite{van2021openasreview}. An initial manual round established the training set, followed by automated runs with different configurations until a predefined stopping rule was reached (100 consecutive irrelevant papers)~\cite{van2023artificialstop}. Papers were retained if at least one round suggested inclusion. This stage reduced the dataset from 3,146 to 215 candidate studies. Replication materials are available in the replication package~\cite{odea2026slrreplication}.

The \textbf{primary screening stage} analysed the titles and abstracts of the 215 papers retained from the preliminary screening. \eoan{In contrast to the preliminary stage, which applied a keyword-based relevance filter, this stage involved a semantic assessment of study relevance based on their meaning and context.} Screening was conducted over three independent rounds applied to the same set of papers: one manually by the first author, and two assisted by a large language model (GPT-4o-mini) using a fixed screening prompt~\cite{thode2025exploring}. The model configuration remained consistent across runs (temperature 0), and the prompt and script are available in the replication package~\cite{odea2026slrreplication}. During this stage, the primary screening inclusion criteria, along with the applicable exclusion criteria (\textit{EC4-EC6}), were applied to determine whether studies addressed energy considerations in microservices within a software engineering or software architectural context. Across rounds, a paper was kept if at least one round suggested inclusion. This process resulted in 117 candidate studies.

The \textbf{secondary screening stage} analysed the introduction and conclusion sections of papers selected after the primary screening to further refine the selection by verifying relevance to the research questions. The screening was performed by one author, with a second author consulted in cases of uncertainty. During this stage, the secondary screening inclusion criteria, along with the applicable exclusion criteria (\textit{EC4, EC5}), were applied based on evidence from the introduction and conclusion sections. As in the primary screening, papers meeting at least one inclusion criterion and none of the exclusion criteria were retained, yielding a final set of 55 primary studies.

The \textbf{final screening stage} involved a full-text analysis of the papers selected in the secondary screening. At this stage, studies were evaluated using a parameter schema derived from the research questions. The parameters are defined in the subsequent RQ-specific subsections (RQ1-RQ3), where they are introduced and applied. The resulting classifications are reported in the corresponding results sections, while the complete parameter definition and per-study assessments are provided in the replication package~\cite{odea2026slrreplication}. Each paper was assessed across several analytical dimensions, including lifecycle considerations, architectural elements, energy measurement methods, analysis tools, metrics, proposed architectural solution, quality trade-offs, and adoption challenges. These parameters were used to quantify the relevance and analytical contribution of each study to the review. Each parameter was scored in the range [0,1], indicating the degree to which the study addressed the corresponding aspect. The results were used to assess relevance and select the studies included in the final synthesis. This stage resulted in a final set of 37 primary studies.

Complementing the database search, a \textbf{snowballing process} was conducted to identify additional relevant studies that may not be retrieved through keyword-based database queries, such as papers using alternative terminology or references outside indexed search fields, following the guidelines proposed by Wohlin~\cite{wohlin2014guidelines}. Both backward and forward snowballing were applied to the papers retained after the final screening. This process identified 2,504 additional candidate papers (1,503 papers backward and 1,001 forward). These papers were subjected to the same screening protocol used for the database search. After screening, only three additional studies were retained, suggesting that the database search had already captured most of the relevant literature.
\section{Energy Efficiency in Microservice Architectures (RQ1)}\label{sc:rq1}

This section addresses \textit{RQ1: Where is energy efficiency considered in microservice architecture research?} It examines two dimensions: (i) the stages of the software lifecycle at which energy efficiency is addressed, and (ii) the architectural elements and relationships through which energy consumption is measured and analysed.

\subsection{RQ1.1: At which stages of the software lifecycle is energy efficiency considered in microservices?}\label{subsc:rq1.1}

This sub-question analyses the stages of the software lifecycle at which energy efficiency becomes a stated concern. This enables us to assess whether energy efficiency is treated as a lifecycle-spanning concern or primarily addressed during specific phases. To support this analysis, we define three parameters: \rqparambf{rq1.1.p1} captures the lifecycle stage at which energy-related concerns are analysed, measured or reported in the literature; \rqparambf{rq1.1.p2} represents the stage at which energy-aware mechanisms are actively designed or implemented within the system; and \rqparambf{rq1.1.p3} characterises how energy efficiency is incorporated through architectural or runtime mechanisms. 

\begin{figure}[t]
    \centering
    \includegraphics[width=.8\linewidth]{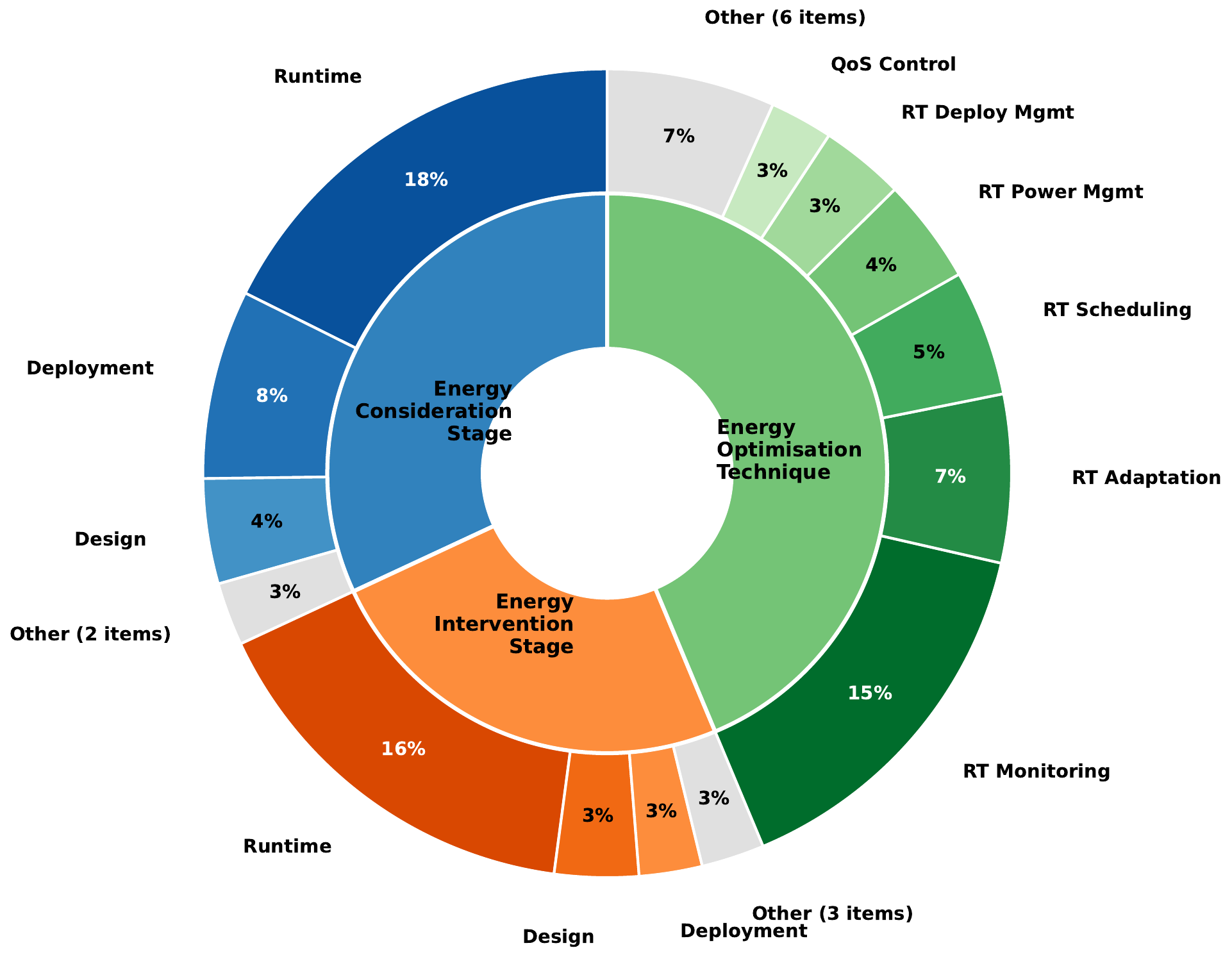}
    \caption{Combined distribution of \rqparam{rq1.1.p1}, \rqparam{rq1.1.p2}, and \rqparam{rq1.1.p3}. Percentages are calculated within each parameter category. In all nested donut charts, the inner ring shows parameter-level distributions, while the outer ring shows the most frequent values, with less frequent categories aggregated. Equal outer-ring percentages may be truncated for clarity.}
    \label{fig:rq1.1_nested_combined}
\end{figure}

Figure~\ref{fig:rq1.1_nested_combined} provides an overview of how energy efficiency is considered across the \eoan{three dimensions.} \eoan{These results show that energy efficiency in microservice architectures is predominantly treated as a runtime concern, with strong alignment across where it is analysed, where it is implemented, and how it is operationalised. Across both \rqparam{rq1.1.p1} and \rqparam{rq1.1.p2}, energy efficiency is primarily concentrated at runtime, where over half of the studies report energy-related concerns and nearly two-thirds introduce energy-aware mechanisms, with deployment emerging as a secondary stage and comparatively limited consideration during design. A similar pattern is observed for \rqparam{rq1.1.p3}, which is strongly dominated by runtime mechanisms such as monitoring, adaptation, and scheduling, with runtime monitoring alone accounting for the largest share of reported techniques, while non-runtime approaches remain sparse.}

\eoan{This alignment reveals a consistent pattern: energy efficiency is primarily treated as a runtime optimisation concern rather than a design-time architectural objective. Energy-related analysis and intervention tend to occur after system structure, service boundaries, and communication patterns have already been established. As a result, optimisation efforts focus on adjusting operational behaviour rather than influencing foundational architectural decisions.}

\eoan{This tendency reflects the inherent characteristics of microservice architectures, where decentralised control, dynamic scaling, and runtime orchestration provide natural leverage points for energy optimisation. However, it also highlights a limited emphasis on embedding energy efficiency into early design activities, such as service decomposition, data management strategies, or inter-service communication design. Consequently, energy efficiency is mainly addressed in a reactive manner, constraining opportunities for more fundamental, design-driven improvements.}


\paragraph{\textbf{Cross-parameters Analysis}}
\begin{table}[b]
\centering
\small
\begin{tabular}{p{5cm} | l l}
\hline
\multirow{2}{*}{\textbf{Integration Approach (RQ1.1.P3)}} &
\multicolumn{2}{c}{\textbf{Lifecycle Stage (RQ1.1.P1)}} \\
\cline{2-3}
&
\textbf{Design} &
\textbf{Runtime} \\
\hline

Design Time Analysis &
~\cite{j_a_larracoechea_radiance_nodate} &
-- \\

Design Time Constraints &
~\cite{agos_jawaddi_sn_analyzing_nodate} &
-- \\

Placement Optimization &
-- &
~\cite{k_afachao_efficient_nodate} \\

QoS Driven Control &
-- &
~\cite{h_h_a_valera_draceo_nodate, c_song_service_nodate, n_toosi_greenfog_nodate} \\

Runtime Adaptation &
-- &
~\cite{xu_m_energy_nodate, adeppady_m_dynamic_nodate, wang_l_energy-delay-aware_nodate, xu_m_energy_nodate-1, m_xu_ibrownout_nodate} \\

Runtime Autoscaling &
-- &
~\cite{gunasekaran_fifer_nodate} \\

\raggedright Runtime Deployment Management &
~\cite{j_von_kistowski_teastore_nodate} &
~\cite{h_humberto_alvarez_valera_pisco_nodate} \\

Runtime Monitoring &
~\cite{dinga_empirical_nodate} &
~\cite{z_xiang_x-man_nodate, saboor_a_containerized_nodate, vitali_towards_nodate, bhasi_kraken_nodate, valera_energy_nodate, z_bellal_gas_nodate, v_berry_is_nodate, brondolin_black-box_nodate} \\
\hline
\end{tabular}
\caption{Mapping of \rqparam{rq1.1.p3} to \rqparam{rq1.1.p1}. Only studies reporting both dimensions are shown.}
\label{tab:rq1.1.p1_p3_mapping}
\end{table}

Table~\ref{tab:rq1.1.p1_p3_mapping} maps \rqparam{rq1.1.p3} to \rqparam{rq1.1.p1}, revealing that design-stage awareness is associated only with a small subset of approaches, primarily static mechanisms such as design-time analysis and constraint specification. In contrast, the majority of integration approaches appear exclusively in conjunction with runtime-stage awareness, including adaptation, autoscaling, QoS-driven control, and monitoring. Runtime deployment management is the only integration approach appearing in both design and runtime stages, suggesting that early-stage activities rarely constitute standalone integration strategies. Instead, they typically prepare or configure mechanisms that are executed once the system is operational.

Table~\ref{tab:rq1.1.p2_p3_mapping} maps \rqparam{rq1.1.p2} to \rqparam{rq1.1.p3}, showing that runtime-oriented approaches are predominantly integrated at runtime, with limited preparation at earlier integration points such as design-time or deployment-time. While a small number of approaches involve non-runtime integration points, these cases still culminate in runtime execution, indicating that early-stage activities primarily serve as enabling steps rather than as locations for energy efficiency action. No integration approach is associated exclusively with design-time or analysis-stage awareness without a corresponding runtime mechanism. Design-time and analysis-stage considerations appear infrequently and show weaker cross-parameter coupling, typically aligning with static analysis, modelling, or constraint specification rather than continuous operational control. These patterns indicate that energy efficiency is operationalised primarily through mechanisms that depend on live system feedback, rather than being embedded as a lifecycle-spanning architectural concern.

\begin{table}[t]
\centering
\small
\renewcommand{\arraystretch}{1.2}
\begin{tabular}{p{3.5cm} | l l l l}
\hline
\multirow{2}{*}{\raggedright\textbf{Integration Approach (P3)}} &
\multicolumn{4}{c}{\textbf{Energy integration point (P2)}} \\
\cline{2-5}
&
\textbf{Configuration} &
\textbf{Deployment} &
\textbf{Design} &
\textbf{Runtime} \\
\hline

Design Time Analysis &
-- &
-- &
\cite{v_berry_is_nodate} &
-- \\

\raggedright Design Time Constraints &
-- &
-- &
\cite{cortellessa_v_exploring_nodate} &
-- \\

\raggedright Placement Optimization &
-- &
-- &
-- &
\cite{w_villegas-ch_adaptive_nodate} \\

\raggedright QoS Driven Control &
\cite{xu_m_energy_nodate} &
-- &
-- &
\cite{n_toosi_greenfog_nodate, khairy_simr_nodate} \\

\raggedright Runtime Adaptation &
-- &
-- &
-- &
\cite{gunasekaran_fifer_nodate, j_gedeon_microservice_nodate, y_huang_satedge_nodate, vitali_towards_nodate, brondolin_black-box_nodate} \\

\raggedright Runtime Autoscaling &
-- &
-- &
-- &
\cite{adeppady_m_dynamic_nodate} \\

\raggedright Runtime Deployment Management &
-- &
\cite{agos_jawaddi_sn_analyzing_nodate} &
-- &
\cite{bhasi_kraken_nodate} \\

\raggedright Runtime Monitoring &
-- &
-- &
-- &
\cite{i_f_model-driven_nodate, j_von_kistowski_teastore_nodate, valera_energy_nodate, z_bellal_gas_nodate, e_ahvar_deca_nodate, a_mokhtari_towards_nodate, g_h_prathama_green_nodate} \\

\hline
\end{tabular}
\caption{Mapping of \textit{integration approach (RQ1.1.P3)} to \textit{energy integration point (RQ1.1.P2)}. Only studies explicitly reporting both parameters are shown.}
\label{tab:rq1.1.p2_p3_mapping}
\end{table}

\begin{framed}
RQ1.1 - Energy efficiency in microservices is predominantly treated as a runtime concern, with limited integration into early lifecycle stages and minimal evidence of design-driven optimisation strategies.
\end{framed}


\begin{figure}[b]
    \centering
    \includegraphics[width=.8\linewidth]{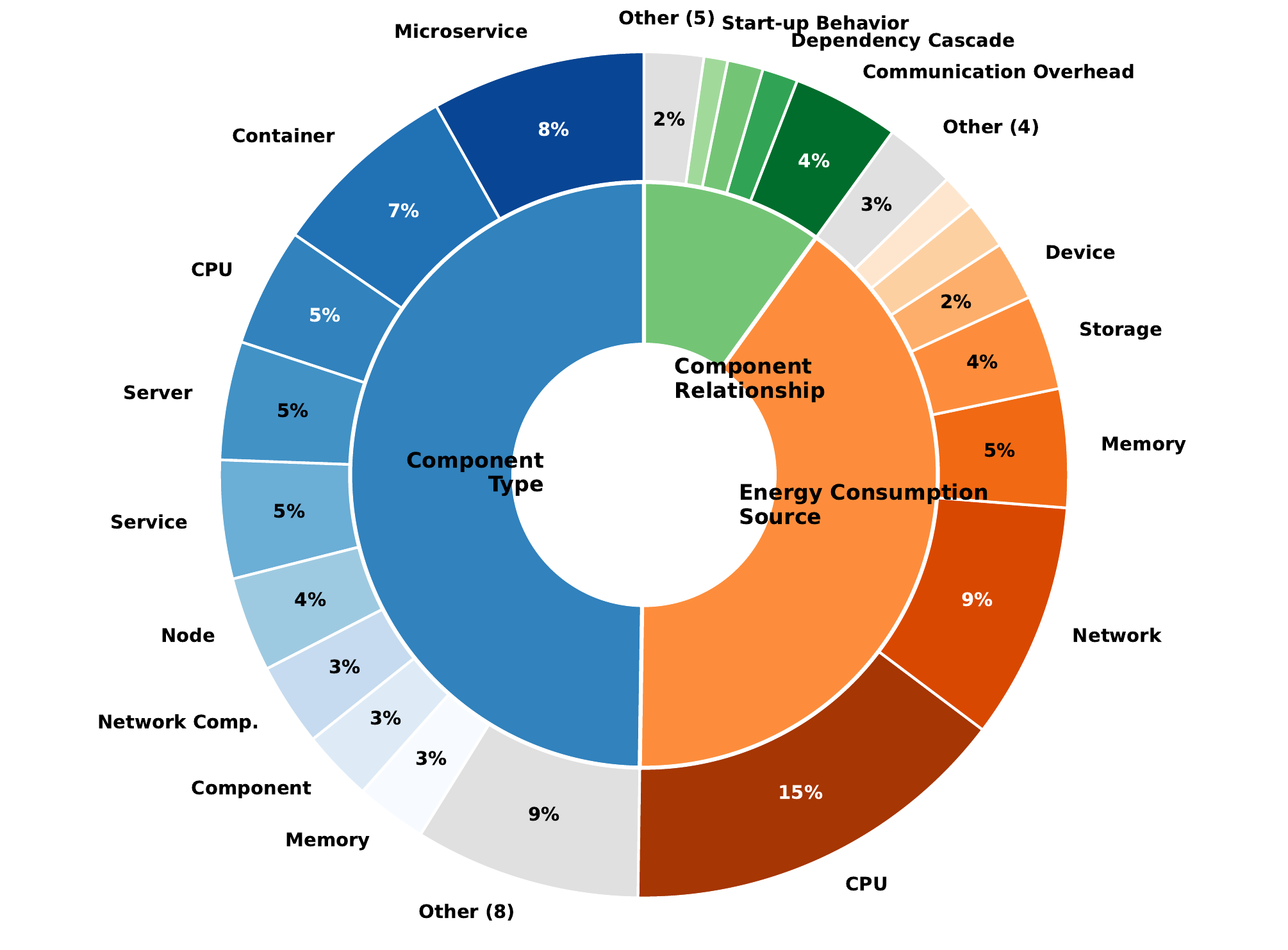}
    \caption{Combined distribution of \rqparam{rq1.2.p1}, \rqparam{rq1.2.p2} and \rqparam{rq1.2.p3}. Percentages calculated relative to the total within each parameter category. Equal outer-ring percentages are truncated for clarity.}
    \label{fig:rq1.2_nested_combined}
\end{figure}

\subsection{RQ1.2: Which microservice architectural elements are primarily considered when measuring energy consumption?}\label{subsc:rq1.2}
\eoan{This question examines how energy consumption is measured across different parts of microservice architectures. It focuses on which architectural elements are analysed (e.g. individual microservices, containers, servers), what system activities contribute to energy use, and how relationships between components are considered when reasoning about energy efficiency. To address \textit{RQ1.2}, we define three parameters: \rqparambf{rq1.2.p1}, which captures where energy consumption is measured within the system; \rqparambf{rq1.2.p2}, which identifies the types of activity contributing to energy use; and \rqparambf{rq1.2.p3}, which describes how interactions between components are considered. As primary studies use inconsistent terminology and operate at different levels, components are grouped based on where energy is measured, rather than a fixed architectural classification.}

\eoan{Figure~\ref{fig:rq1.2_nested_combined} summarises the distribution of coded instances across the three parameters. Energy consumption in microservice architectures is analysed across multiple abstraction levels, but with a clear emphasis on individual components and low-level resource usage, while interactions between components remain comparatively underexplored. For \rqparam{rq1.2.p1}, energy measurement is most often attributed to specific architectural elements, with individual microservices and containers most frequently considered, alongside infrastructure components such as CPUs and servers. This indicates that energy analysis is distributed across both application- and infrastructure-level elements, rather than being consistently anchored to a single architectural layer.
For \rqparam{rq1.2.p2}, CPU-related consumption dominates, followed by network activity, while other hardware contributors receive comparatively less attention. This reflects a strong bias toward compute- and communication-centric interpretations of energy-use, with comparatively limited consideration of other system resources. In contrast, \rqparam{rq1.2.p3} is less frequently addressed and is mainly limited to communication overhead. This imbalance highlights that, despite the inherently distributed nature of microservices, energy efficiency is rarely analysed in terms of inter-component dynamics, and is instead mainly treated as a property of individual components.}

\eoan{Overall, these observations suggest that current research tends to focus on where energy is consumed within isolated components, rather than how architectural structure and interactions shape overall energy behaviour.}

\begin{table}[b]
\centering
\small
\renewcommand{\arraystretch}{1.2}
\begin{tabular}{l | p{3cm} p{1.5cm} p{1.5cm} p{1.5cm} l}
\hline
\multirow{2}{*}{\raggedright\textbf{Component Type (RQ1.2.P1)}} &
\multicolumn{5}{c}{\textbf{Energy consumption source (RQ1.2.P2)}} \\
\cline{2-6}
&
\textbf{CPU} &
\textbf{Device} &
\textbf{Idle Power} &
\textbf{Memory} &
\textbf{Power Supply} \\
\hline

Container &
\cite{s_r_chaudhry_improved_nodate, z_xiang_x-man_nodate, wang_l_energy-delay-aware_nodate, k_afachao_efficient_nodate, c_song_service_nodate, n_toosi_greenfog_nodate, j_von_kistowski_teastore_nodate, agos_jawaddi_sn_analyzing_nodate, m_vitali_enriching_nodate, brondolin_black-box_nodate, c_courageux-sudan_studying_nodate, m_s_floroiu_anomaly_nodate, antoniou_agile_nodate} &
\cite{m_xu_ibrownout_nodate} &
-- &
\cite{z_bellal_gas_nodate} &
-- \\

CPU &
\cite{khairy_simr_nodate, h_humberto_alvarez_valera_pisco_nodate, e_ahvar_deca_nodate} &
\cite{xu_m_energy_nodate-1} &
\cite{saboor_a_containerized_nodate} &
-- &
-- \\

Device &
\cite{gunasekaran_fifer_nodate} &
\cite{y_huang_satedge_nodate} &
-- &
-- &
-- \\

Gateway &
\cite{a_mokhtari_towards_nodate} &
-- &
-- &
-- &
-- \\

Memory &
\cite{vitali_towards_nodate} &
-- &
-- &
\cite{calagna_a_enabling_nodate} &
-- \\

Microservice &
\cite{adeppady_m_dynamic_nodate, w_villegas-ch_adaptive_nodate, i_f_model-driven_nodate, valera_energy_nodate, g_h_prathama_green_nodate} &
-- &
-- &
-- &
-- \\

Network Component &
-- &
-- &
-- &
-- &
-- \\

Node &
\cite{bhasi_kraken_nodate} &
-- &
-- &
-- &
-- \\

Service &
\cite{dinga_empirical_nodate} &
-- &
-- &
-- &
\cite{cortellessa_v_exploring_nodate} \\

\hline
\end{tabular}
\caption{Mapping of \textit{component type (RQ1.2.P1)} to \textit{energy consumption source (RQ1.2.P2)}. Only studies explicitly reporting both parameters are shown.}
\label{tab:rq1.2.p1_p2_mapping}
\end{table}

\paragraph{\textbf{Cross-parameters Analysis}}
Table~\ref{tab:rq1.2.p1_p2_mapping} maps \rqparam{rq1.2.p1} to \rqparam{rq1.2.p2}, showing that CPU-related energy consumption is reported for nearly all component categories, making it the most frequently considered energy source across a wide range of architectural component types. Other energy consumption sources appear only sporadically and are limited to a small subset of components. Containers emerge as the most frequently reported component type for CPU-related energy measurement, reflecting their role as a common execution and deployment unit. By contrast, non-CPU energy sources, such as device-level energy, idle power, memory, and power supply, are reported in only a small number of isolated pairings and are not consistently associated with any single component type. No architectural component is predominantly associated with these sources, and they do not form coherent measurement clusters within the dataset.

\begin{table}[t!]
\centering
\small
\renewcommand{\arraystretch}{1.2}
\begin{tabular}{l | p{1.5cm} p{1.5cm} p{1.5cm} p{1.5cm} l}
\hline
\multirow{2}{*}{\raggedright\textbf{Component Relationship (RQ1.2.P3)}} &
\multicolumn{5}{c}{\textbf{Energy consumption source (RQ1.2.P2)}} \\
\cline{2-6}
&
\textbf{CPU} &
\textbf{Device} &
\textbf{Idle Power} &
\textbf{Memory} &
\textbf{Power Supply} \\
\hline

Dependency Cascade &
\cite{j_gedeon_microservice_nodate} &
\cite{y_huang_satedge_nodate} &
-- &
\cite{e_ahvar_deca_nodate} &
-- \\

Placement Dependency &
\cite{wang_l_energy-delay-aware_nodate, j_a_larracoechea_radiance_nodate} &
-- &
-- &
-- &
\cite{h_humberto_alvarez_valera_pisco_nodate} \\

Reconfiguration &
-- &
\cite{saboor_a_containerized_nodate} &
-- &
-- &
-- \\

Redundancy Replication &
\cite{khairy_simr_nodate} &
-- &
-- &
-- &
-- \\

Request Batching &
\cite{j_von_kistowski_teastore_nodate} &
-- &
-- &
-- &
-- \\

Service Deactivation &
-- &
-- &
\cite{i_f_model-driven_nodate} &
-- &
-- \\

Shared Resource Contention &
\cite{g_h_prathama_green_nodate} &
-- &
-- &
-- &
-- \\

Start Up Behavior &
\cite{yu_y_joint_nodate} &
-- &
-- &
-- &
-- \\

\hline
\end{tabular}
\caption{Mapping of \textit{component relationship (RQ1.2.P3)} to \textit{energy consumption source (RQ1.2.P2)}. Only studies explicitly reporting both parameters are shown.}
\label{tab:rq1.2.p2_p3_mapping}
\end{table}


Table~\ref{tab:rq1.2.p2_p3_mapping} maps \rqparam{rq1.2.p2} to \rqparam{rq1.2.p3}, showing that CPU-related energy consumption co-occurs with a wide range of component relationship types, spanning multiple interaction- and dependency-level relationships. Other sources appear less frequently and are associated with a narrower subset of relationship types. In contrast, non-CPU energy sources exhibit far more constrained associations with component relationships. Device-related energy appears only in relation to dependency cascades and reconfiguration. Idle power is reported exclusively in the context of service deactivation. Memory and power supply energy sources are each associated with a single relationship type. These patterns indicate that no component relationship is consistently associated with a specific energy consumption source across the literature. These narrow pairings suggest that non-CPU sources are considered only in highly specific architectural scenarios, rather than serving as general explanatory bases for energy behaviour.


\begin{framed} 
RQ1.2 - The reviewed literature most commonly attributes or analyses energy consumption at the abstraction levels of microservices and containers, alongside host-level compute resources such as CPUs and servers. Energy use is mainly attributed to CPU activity and, to a lesser extent, network-related costs, while other sources such as memory, storage, and idle power receive limited attention. Architectural impacts on energy efficiency are mainly analysed in terms of communication overhead between components, with more complex interaction effects addressed far less frequently.
\end{framed}

\subsection{Answer to the Research Question (RQ1)}

This section synthesises the findings of \textit{RQ1.1-RQ1.2} to answer \textit{RQ1: Where is energy efficiency considered in microservice architecture research?} 
To answer this question, we examine where lifecycle-stage awareness of energy efficiency intersects with specific architectural elements. This allows us to identify not only when energy efficiency is considered, but also which components and energy sources are associated with these considerations across the lifecycle.


\begin{table}[b]
\centering
\small
\renewcommand{\arraystretch}{1.2}
\begin{tabular}{l | l l}
\hline
\multirow{2}{*}{\textbf{Component Type (RQ1.2.P1)}} &
\multicolumn{2}{c}{\textbf{Lifecycle Stage (RQ1.1.P1)}} \\
\cline{2-3}
&
\textbf{Design} &
\textbf{Runtime} \\
\hline

Container &
~\cite{antoniou_agile_nodate} &
~\cite{z_xiang_x-man_nodate, xu_m_energy_nodate-1, khairy_simr_nodate, j_von_kistowski_teastore_nodate, cortellessa_v_exploring_nodate, h_humberto_alvarez_valera_pisco_nodate, calagna_a_enabling_nodate, g_h_prathama_green_nodate} \\

CPU &
~\cite{valera_energy_nodate} &
~\cite{saboor_a_containerized_nodate, c_song_service_nodate, vitali_towards_nodate, m_vitali_enriching_nodate} \\

Device &
-- &
~\cite{j_gedeon_microservice_nodate} \\

Gateway &
-- &
~\cite{dinga_empirical_nodate} \\

Memory &
-- &
~\cite{j_a_larracoechea_radiance_nodate} \\

Microservice &
~\cite{a_mokhtari_towards_nodate} &
~\cite{s_r_chaudhry_improved_nodate, y_huang_satedge_nodate, e_ahvar_deca_nodate} \\

Service &
~\cite{bhasi_kraken_nodate} &
-- \\

\hline
\end{tabular}
\caption{Mapping of \textit{component type (RQ1.2.P1)} to \textit{lifecycle stage energy awareness (RQ1.1.P1)}. Only studies reporting both dimensions are shown.}
\label{tab:rq1.1.p1_rq1.2.p1_mapping}
\end{table}
\begin{table}[t]
\centering
\small
\renewcommand{\arraystretch}{1.2}

\centering
\small
\renewcommand{\arraystretch}{1.2}
\begin{tabular}{l | l l l l}
\hline
\multirow{2}{*}{\textbf{Energy Consumption Source (RQ1.2.P2)}} &
\multicolumn{4}{c}{\textbf{Energy Awareness Stage (RQ1.1.P2)}} \\
\cline{2-5}
&
\textbf{Configuration} &
\textbf{Deployment} &
\textbf{Design} &
\textbf{Runtime} \\
\hline

CPU &
~\cite{s_r_chaudhry_improved_nodate} &
-- &
~\cite{z_bellal_gas_nodate, dinga_empirical_nodate} &
~\cite{yu_y_joint_nodate, wang_l_energy-delay-aware_nodate, m_xu_ibrownout_nodate, j_von_kistowski_teastore_nodate, h_humberto_alvarez_valera_pisco_nodate, bhasi_kraken_nodate, calagna_a_enabling_nodate, j_a_larracoechea_radiance_nodate, m_s_floroiu_anomaly_nodate, g_h_prathama_green_nodate} \\

Device &
-- &
-- &
-- &
~\cite{i_f_model-driven_nodate, agos_jawaddi_sn_analyzing_nodate} \\

IdlePower &
-- &
-- &
-- &
~\cite{n_toosi_greenfog_nodate} \\

Memory &
-- &
-- &
-- &
~\cite{m_vitali_enriching_nodate, v_berry_is_nodate} \\

PowerSupply &
-- &
~\cite{valera_energy_nodate} &
-- &
-- \\

\hline
\end{tabular}
\caption{Mapping of \textit{energy consumption source (RQ1.2.P2)} to \textit{energy integration point (RQ1.1.P2)}. Only studies reporting both dimensions are shown.}
\label{tab:rq1.1.p2_rq1.2.p2_mapping}
\end{table}

Table~\ref{tab:rq1.1.p1_rq1.2.p1_mapping} maps \rqparam{rq1.1.p1} to \rqparam{rq1.2.p1}, showing that energy efficiency considerations for most architectural component types are concentrated at runtime. Across mapped studies, architectural components are predominantly associated with runtime-stage awareness, while design-time consideration appears for a limited subset of components. No component type exhibits consistent energy-related consideration across both early and late lifecycle stages, indicating that architectural elements are typically examined from an energy perspective only once systems are operational. Table~\ref{tab:rq1.1.p2_rq1.2.p2_mapping} maps \rqparam{rq1.1.p2} to \rqparam{rq1.2.p2}, which further refines this view by showing that CPU-related energy is the only consumption source considered across multiple lifecycle stages. All other consumption sources are addressed exclusively at runtime, with only isolated exceptions. This indicates that early lifecycle energy awareness is narrowly scoped and largely limited to processing-related activity, while non-CPU energy sources are considered only during system operation.

\begin{framed} 
RQ1 - Energy efficiency in microservice architectures is concentrated at runtime, where energy consumption is attributed to a mix of application-level abstractions (microservices and containers) and host-level compute resources. Modelling is heavily centred on CPU activity, while other sources and cross-service interaction effects receive comparatively limited attention.
\end{framed}

\section{Energy Consumption Measurement in Microservices (RQ2)}\label{sc:rq2}

This section addresses \textit{RQ2: How is energy consumption measured in microservices?}, which characterises how energy measurement is considered and assessed in the literature. Unlike \textit{RQ1}, which considers lifecycle stages and architectural levels of energy efficiency, \textit{RQ2} focuses on the practices used to quantify energy consumption. The analysis covers three aspects: (i) measurement methods; (ii) analysis tools, their deployment, and granularity; and (iii) reported metrics and how energy-related performance is quantified, normalised, and evaluated.

\subsection{RQ2.1: What methods are used to measure energy consumption in microservices?}\label{subsc:rq2.1}
This question aims to characterise the technical approaches through which energy data is obtained and analysed in empirical studies. It considers the system levels at which energy is measured, the techniques used to quantify consumption, and the data collection and implementation approaches through which measurement is realised. To support this analysis, we define three parameters: \rqparambf{rq2.1.p1} represents the architectural level at which energy consumption is measured or attributed (e.g., services, containers, hosts, etc.); \rqparambf{rq2.1.p2} describes the approach used to measure energy consumption; and \rqparambf{rq2.1.p3} refers to how energy data is gathered during system execution or experimentation.

\begin{figure}
    \centering
    \includegraphics[width=.7\linewidth]{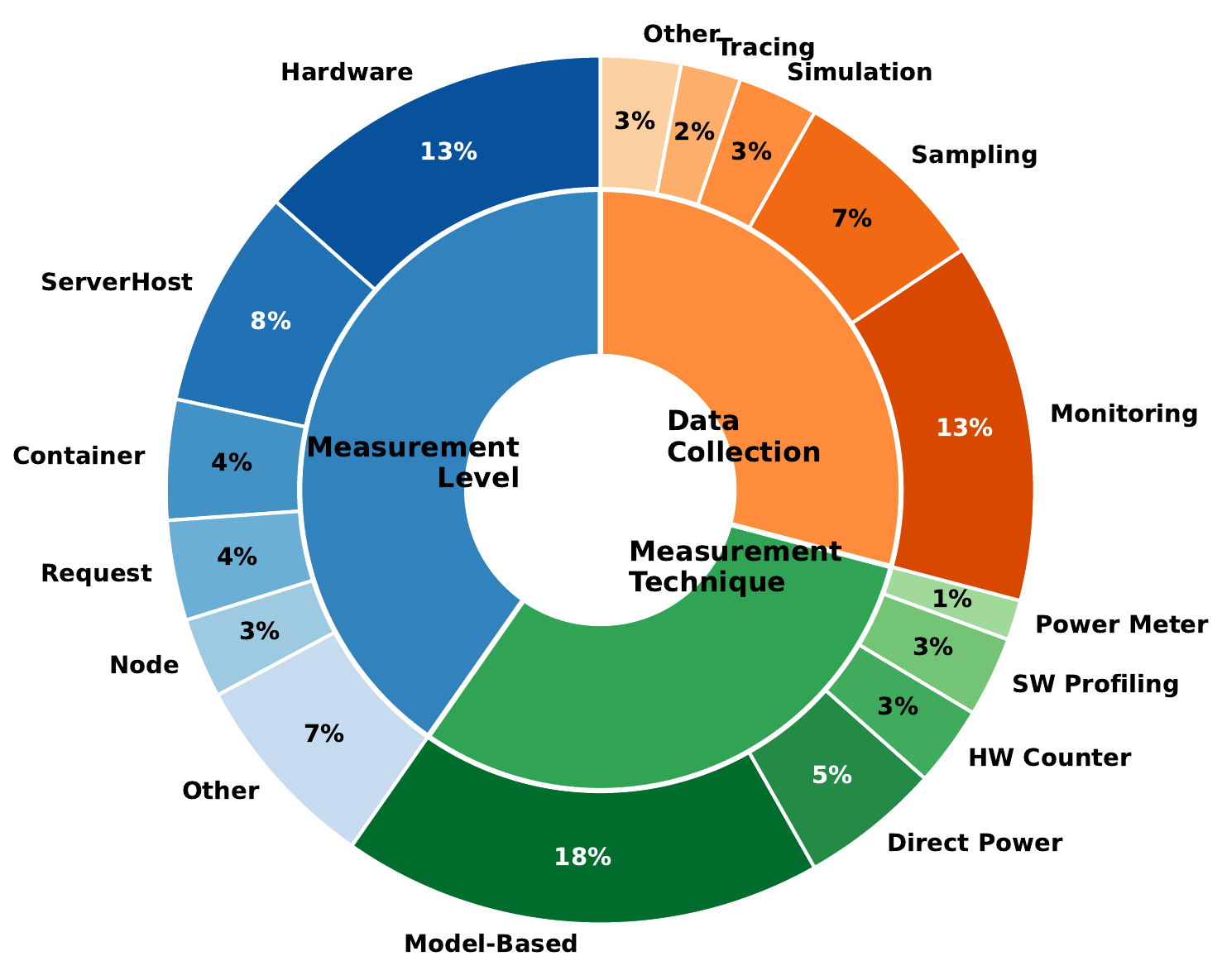}
    \caption{Combined distribution of \rqparam{rq2.1.p1}, \rqparam{rq2.1.p2} and \rqparam{rq2.1.p3}. Percentages calculated relative to the total within each parameter category.}
    \label{fig:rq2.1_nested_combined}
\end{figure}

Figure~\ref{fig:rq2.1_nested_combined} combines the three parameters within a single visual representation. \eoan{Energy measurement in microservice systems is mainly conducted at infrastructure-adjacent levels and relies heavily on indirect estimation techniques, with data typically collected through continuous runtime observation rather than controlled experimentation.} Regarding \rqparam{rq2.1.p1}, measurement practices are strongly concentrated at infrastructure-adjacent levels, with hardware- and server-level measurements accounting for over half of all reported cases, while container-, request, and node-level measurements appear less frequently. \eoan{This indicates that energy is most often captured at coarse-grained system boundaries rather than being directly attributed to individual services, reflecting the difficulty of isolating energy consumption within fine-grained components.}

Across \rqparam{rq2.1.p2}, model-based estimation dominates the reported measurement techniques, while direct power measurement, hardware counters, software profiling, and physical power meters appear in comparatively fewer studies. \eoan{In these approaches, energy consumption is inferred from system metrics such as CPU utilisation or resource usage using predefined models, indicating a preference for scalable estimation over precise instrumentation.} With respect to \rqparam{rq2.1.p3}, it is most commonly realised through continuous monitoring and sampling, together accounting for over two-thirds of reported approaches, while simulation- and tracing-based collection methods appear comparatively infrequently. \eoan{This reflects a tendency to observe energy behaviour during live system execution rather than through controlled or fine-grained experimental methods.}

\paragraph{\textbf{Cross-parameters Analysis}}

\begin{table}[t]
\centering
\small
\renewcommand{\arraystretch}{1.2}
\begin{tabular}{l | p{2cm} p{3cm} p{2cm} p{1.5cm}}
\hline
\multirow{2}{*}{\textbf{Measurement Level (RQ2.1 -- P1)}} &
\multicolumn{4}{c}{\textbf{Measurement Technique (RQ2.1 -- P2)}} \\
\cline{2-5}
&
\textbf{Hardware Counter API} &
\raggedright \textbf{Model-Based Estimation} &
\textbf{Physical Power Meter} &
\textbf{Software Profiling} \\
\hline

Cluster / Datacenter &
-- &
\cite{z_xiang_x-man_nodate, xu_m_energy_nodate-1} &
-- &
-- \\

Component &
-- &
\cite{n_schmitt_online_nodate} &
-- &
-- \\

Container / Pod &
-- &
\cite{vitali_towards_nodate, a_mokhtari_towards_nodate} &
-- &
-- \\

Function &
-- &
\cite{m_s_floroiu_anomaly_nodate} &
-- &
-- \\

Hardware &
-- &
\cite{w_villegas-ch_adaptive_nodate, saboor_a_containerized_nodate, khairy_simr_nodate, m_xu_ibrownout_nodate, agos_jawaddi_sn_analyzing_nodate, cortellessa_v_exploring_nodate, calagna_a_enabling_nodate, j_a_larracoechea_radiance_nodate} &
\cite{c_courageux-sudan_studying_nodate} &
-- \\

Node / Device &
-- &
\cite{i_f_model-driven_nodate, c_song_service_nodate} &
-- &
-- \\

OS &
\cite{k_afachao_efficient_nodate} &
-- &
-- &
-- \\

Request &
\cite{v_berry_is_nodate} &
\cite{wang_l_energy-delay-aware_nodate} &
-- &
\cite{g_h_prathama_green_nodate} \\

Server / Host &
\cite{bhasi_kraken_nodate, antoniou_agile_nodate} &
\cite{yu_y_joint_nodate, n_toosi_greenfog_nodate, h_humberto_alvarez_valera_pisco_nodate, z_bellal_gas_nodate, e_ahvar_deca_nodate} &
-- &
-- \\

\hline
\end{tabular}
\caption{Mapping of measurement levels (RQ2.1.P1) to measurement techniques (RQ2.1.P2). Only studies reporting both dimensions are shown.}
\label{tab:rq2.1.p1_p2_mapping}

\end{table}

The mapping between \rqparam{rq2.1.p1} and \rqparam{rq2.1.p2} reveals a clear concentration of model-based estimation dominates across abstraction levels. As shown in Table~\ref{tab:rq2.1.p1_p2_mapping}, estimation approaches, where energy consumption is inferred from utilisation metrics or performance counters, are reported at nearly all levels of measurement. In contrast, hardware-based techniques, such as counter APIs and physical power meters, which obtain readings directly from instrumentation interfaces, are largely confined to lower-level measurement contexts. Software profiling appears only once and is limited to request-level analysis~\cite{g_h_prathama_green_nodate}.

\begin{table*}[t]
\centering
\small
\renewcommand{\arraystretch}{1.2}

\centering
\small
\renewcommand{\arraystretch}{1.2}
\begin{tabular}{l | p{2cm} p{3.4cm} p{2cm} p{1.5cm}}
\hline
\multirow{2}{*}{\textbf{Data Collection Method (RQ2.1.P3)}} &
\multicolumn{4}{c}{\textbf{Measurement Technique (RQ2.1.P2)}} \\
\cline{2-5}
&
\textbf{Hardware Counter API} &
\textbf{Model-Based Estimation} &
\textbf{Physical Power Meter} &
\textbf{Software Profiling} \\
\hline

APIReading &
\cite{k_afachao_efficient_nodate} &
-- &
-- &
-- \\

Logging &
-- &
\cite{wang_l_energy-delay-aware_nodate} &
-- &
-- \\

Monitoring &
\cite{bhasi_kraken_nodate, antoniou_agile_nodate} &
\cite{w_villegas-ch_adaptive_nodate, m_xu_ibrownout_nodate, agos_jawaddi_sn_analyzing_nodate, h_humberto_alvarez_valera_pisco_nodate, z_bellal_gas_nodate, calagna_a_enabling_nodate, n_schmitt_online_nodate, a_mokhtari_towards_nodate, m_s_floroiu_anomaly_nodate} &
\cite{c_courageux-sudan_studying_nodate} &
\cite{g_h_prathama_green_nodate} \\

Sampling &
\cite{v_berry_is_nodate} &
\cite{z_xiang_x-man_nodate, j_a_larracoechea_radiance_nodate} &
-- &
-- \\

Simulation &
-- &
\cite{c_song_service_nodate, n_toosi_greenfog_nodate, dinga_empirical_nodate} &
-- &
-- \\

Tracing Profiling &
-- &
\cite{cortellessa_v_exploring_nodate} &
-- &
-- \\

\hline
\end{tabular}
\caption{Mapping of data collection methods (RQ2.1.P3) to measurement techniques (RQ2.1.P2). Only studies reporting both parameters are shown.}
\label{tab:rq2.1.p2_p3_mapping}
\end{table*}

Table~\ref{tab:rq2.1.p2_p3_mapping} maps \rqparam{rq2.1.p2} to \rqparam{rq2.1.p3}, showing how different techniques are associated with specific approaches to collecting energy-related data. Across the reviewed studies, monitoring-based data collection is reported most frequently, particularly in combination with model-based estimation, and occurs in conjunction with all reported measurement techniques. This indicates that continuous or periodic observation of system behaviour is the primary means by which energy-related data is obtained, regardless of the underlying measurement approach. Other data collection methods are less frequent and tend to align with specific measurement strategies. Sampling and simulation are used primarily with model-based estimation, reflecting analytical or experimental evaluation settings rather than continuous runtime observation. API-based readings and tracing appear only in isolated cases, including limited use of fine-grained or program-level instrumentation. Physical power meters are rare and embedded within monitoring setups, suggesting that direct hardware measurement is typically integrated into broader monitoring infrastructures rather than used independently. This analysis shows that energy measurement is largely estimation-driven and monitoring-supported, with direct hardware instrumentation playing a limited and infrastructure-focused role.

\begin{framed} 
RQ2.1 - Energy consumption in microservices is most commonly measured at infrastructure- and container adjacent levels using model-based estimation techniques, supported primarily by monitoring-oriented data collection, while direct measurement approaches such as hardware counters, physical power meters, and software profiling are used less frequently and remain narrowly scoped.
\end{framed}

\begin{figure}[b]
    \centering
    \includegraphics[width=.7\linewidth]{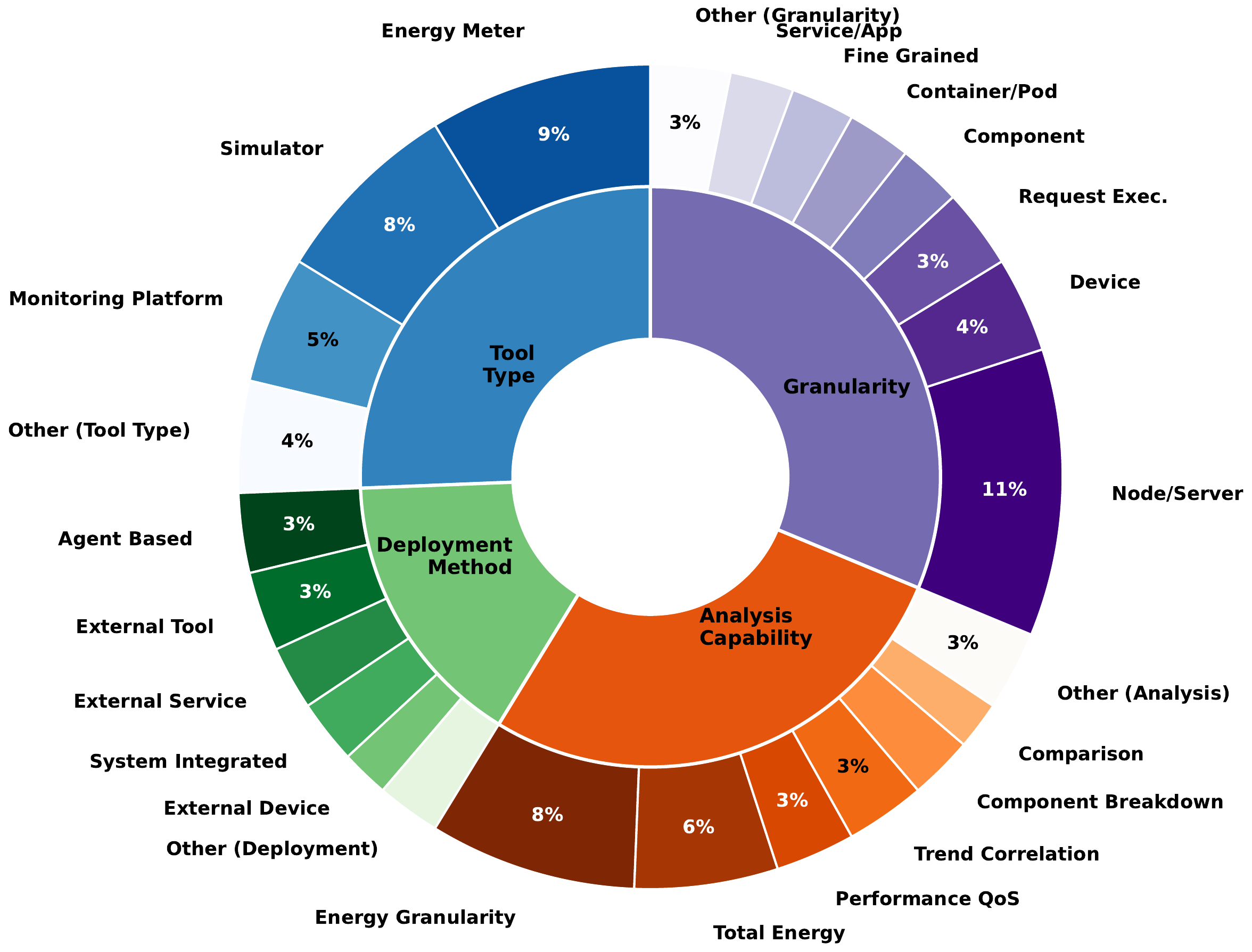}
    \caption{Combined distribution of \rqparam{rq2.2.p1}, \rqparam{rq2.2.p2}, \rqparam{rq2.2.p3}, and \rqparam{rq2.2.p4}. Percentages calculated relative to the total within each parameter category.}
    \label{fig:rq2.2_nested_combined}
\end{figure}

\subsection{RQ2.2: How are tools used for energy consumption analysis in microservices?}\label{subsc:rq2.2}
This question aims to characterise the types of tools employed, how they are deployed within experimental or operational settings, and the forms of analysis and levels of granularity they support. To support this analysis, we define four parameters: \rqparambf{rq2.2.p1}, which categorizes the kinds of tools used for energy consumption analysis; \rqparambf{rq2.2.p2}, which describes how tools are integrated or deployed within the system or experimental setup; \rqparambf{rq2.2.p3}, which captures the types of energy-related insights the tools provide; and \rqparambf{rq2.2.p4}, which indicates the level of detail at which energy consumption is analysed.

Figure~\ref{fig:rq2.2_nested_combined} summarises the characteristics of tools used for energy consumption analysis across the four dimensions: tool type, deployment method, analysis capability and measurement granularity. \eoan{Energy analysis tools in microservice systems are predominantly oriented toward infrastructure-level measurement and runtime integration, with limited support for fine-grained, architecture-aligned analysis.} With respect to \rqparam{rq2.2.p1}, energy meters and simulators are the most frequently reported tool categories, followed by monitoring platforms. \eoan{Representative examples include simulation frameworks such as \textit{CloudSim}, \textit{SimGrid}, and \textit{PISCO}, as well as hardware-level measurement interfaces such as \textit{Turbostat} and \textit{RAPL}.} \eoan{This indicates a reliance on tools that either approximate energy consumption through modelling or capture it at the hardware and system level, rather than directly measuring energy at the level of individual services or interactions.}

In terms of \rqparam{rq2.2.p2}, tools are commonly deployed as agent-based components~\cite{c_song_service_nodate}, external tools or services~\cite{gunasekaran_fifer_nodate}, or system-integrated mechanisms~\cite{j_gedeon_microservice_nodate}. Agent-based approaches typically embed lightweight monitoring components within nodes or containers, enabling continuous collection of resource utilisation metrics during execution, \eoan{reflecting a preference for runtime-integrated monitoring over standalone measurement setups.} Regarding \rqparam{rq2.2.p3}, tool support is primarily focused on reporting total energy consumption and coarse-grained energy profiles, with fewer studies providing capabilities such as trend correlation, comparative analysis, or detailed component breakdown. \eoan{This limits analysis to aggregate observations rather than supporting deeper reasoning about how energy consumption relates to architectural structure or behaviour.}

Finally, \rqparam{rq2.2.p4} shows that measurement granularity is largely coarse-grained, with node- or server-level analysis dominating, while request-level, function-level, and fine-grained service measurements are less supported. \eoan{This reinforces the disconnect between current measurement practices and the fine-grained components and interactions that define microservice architectures.}

\begin{table*}[b]
\centering
\small
\renewcommand{\arraystretch}{1.2}
\begin{tabular}{l | p{1.2cm} p{1.2cm} p{1.3cm} p{1.2cm} p{1.4cm} p{.8cm} p{1.4cm}}
\hline
\multirow{2}{*}{\textbf{Analysis Capability (RQ2.2.P3)}} &
\multicolumn{7}{c}{\textbf{Tool Type (RQ2.2.P1)}} \\
\cline{2-8}
&
\textbf{Estimation Tool} &
\textbf{Energy Meter} &
\textbf{Framework} &
\textbf{Model Checker} &
\textbf{Monitoring Platform} &
\textbf{OS API} &
\textbf{Simulator} \\
\hline

Component Breakdown &
-- &
\cite{z_xiang_x-man_nodate, y_huang_satedge_nodate} &
-- &
-- &
-- &
-- &
\cite{cortellessa_v_exploring_nodate} \\

Energy Granularity &
\cite{w_villegas-ch_adaptive_nodate} &
\cite{v_berry_is_nodate, m_s_floroiu_anomaly_nodate} &
-- &
-- &
\cite{a_mokhtari_towards_nodate, g_h_prathama_green_nodate} &
\cite{k_afachao_efficient_nodate} &
\cite{s_r_chaudhry_improved_nodate, i_f_model-driven_nodate, calagna_a_enabling_nodate, j_a_larracoechea_radiance_nodate} \\

Pattern Detection &
-- &
-- &
-- &
-- &
-- &
-- &
\cite{dinga_empirical_nodate} \\

Performance QoS &
-- &
-- &
\cite{c_song_service_nodate} &
-- &
-- &
-- &
\cite{n_toosi_greenfog_nodate} \\

Sustainability &
\cite{brondolin_black-box_nodate} &
-- &
-- &
-- &
-- &
-- &
-- \\

Total Energy &
-- &
\cite{agos_jawaddi_sn_analyzing_nodate, h_humberto_alvarez_valera_pisco_nodate, bhasi_kraken_nodate} &
-- &
-- &
-- &
-- &
\cite{wang_l_energy-delay-aware_nodate} \\

Validation Verification &
-- &
-- &
-- &
\cite{z_bellal_gas_nodate} &
-- &
-- &
-- \\

\hline
\end{tabular}
\caption{Mapping of analysis capabilities (RQ2.2.P3) to tool types (RQ2.2.P1). Only studies reporting both parameters are shown.}
\label{tab:rq2.2.p1_p3_mapping}
\end{table*}

\paragraph{\textbf{Cross-parameters Analysis}}
Table~\ref{tab:rq2.2.p1_p3_mapping} maps \rqparam{rq2.2.p1} to \rqparam{rq2.2.p3}, showing a clear coupling between tool types and the analysis capabilities they support. Energy meters and simulators are associated with the broadest range of analysis capabilities. In contrast, monitoring platforms and energy estimation tools are linked to a narrower subset of capabilities, primarily focused on energy granularity and sustainability-oriented analysis. More specialised analysis capabilities occur infrequently and exhibit strong tool-specific coupling. Pattern detection and validation or verification capabilities are reported only in isolated cases and are associated exclusively with simulators and model-checking tools, respectively. No tool type supports advanced analysis capabilities across multiple categories, indicating that energy consumption analysis relies on a small set of dominant tool categories for general-purpose analysis, while specialised capabilities are confined to purpose-specific tools.

\begin{table*}[b]
\centering
\small
\renewcommand{\arraystretch}{1.2}
\begin{tabular}{l | p{1.7cm} p{1.1cm} p{1.3cm} p{1.1cm} p{1.5cm} p{1.26cm}}
\hline
\multirow{2}{*}{\textbf{Tool Type (RQ2.2.P1)}} &
\multicolumn{6}{c}{\textbf{Deployment Method (RQ2.2.P2)}} \\
\cline{2-7}
&
\raggedright \textbf{Container Orchestrated} &
\textbf{External Device} &
\textbf{External Service} &
\textbf{External Tool} &
\raggedright \textbf{Local Application} &
\textbf{System Integrated} \\
\hline

Energy Estimation Tool &
-- &
-- &
-- &
\cite{w_villegas-ch_adaptive_nodate} &
-- &
-- \\

Energy Meter &
-- &
\cite{agos_jawaddi_sn_analyzing_nodate, c_courageux-sudan_studying_nodate} &
\cite{h_humberto_alvarez_valera_pisco_nodate, bhasi_kraken_nodate, m_s_floroiu_anomaly_nodate} &
\cite{z_xiang_x-man_nodate, y_huang_satedge_nodate} &
-- &
-- \\

Framework &
\cite{c_song_service_nodate} &
-- &
-- &
-- &
-- &
-- \\

Model Checker &
-- &
-- &
-- &
\cite{z_bellal_gas_nodate} &
-- &
-- \\

OSAPI &
-- &
-- &
-- &
-- &
-- &
\cite{k_afachao_efficient_nodate} \\

Simulator &
-- &
-- &
-- &
\cite{j_a_larracoechea_radiance_nodate} &
\cite{s_r_chaudhry_improved_nodate, cortellessa_v_exploring_nodate} &
-- \\

\hline
\end{tabular}
\caption{Mapping of tool types (RQ2.2.P1) to deployment methods (RQ2.2.P2). Only studies reporting both parameters are shown.}
\label{tab:rq2.2.p1_p2_mapping}
\end{table*}

Table~\ref{tab:rq2.2.p1_p2_mapping} maps \rqparam{rq2.2.p1} to \rqparam{rq2.2.p2} for studies reporting both parameters, showing a strong coupling between tool types and their deployment methods, with most tool categories associated with a single dominant mode of deployment. Energy meters exhibit the greatest deployment variability, appearing as external devices, external services, and external tools, while other tool types are largely confined to a single deployment approach. Several tool categories display exclusive deployment patterns. Framework-based tools appear only in container-orchestrated setups, OS-level APIs are deployed exclusively as system-integrated mechanisms, and both energy estimation tools and model checkers are reported only as external tools. Simulators are primarily implemented as local applications, with one case of being deployed as an external tool. Container-orchestrated and system-integrated deployments occur infrequently overall, and no tool type spans both deeply embedded and fully external deployment modes. This indicates that energy analysis tools are typically deployed using narrowly defined integration strategies rather than flexible or multi-modal deployment approaches.

\begin{framed} 
RQ2.2 - Energy consumption analysis in microservices is primarily supported by a small set of dominant tool types, most notably energy meters and simulators (e.g. Turbostat, RAPL, CloudSim), which are deployed through tool-specific integration strategies and mainly support coarse-grained energy analysis, while fine-grained and specialised analyses remain confined to a limited number of purpose-built tools.
\end{framed}

\subsection{RQ2.3: What energy efficiency metrics are commonly used to evaluate microservices?}\label{subsc:rq2.3}
This question aims to characterise the types of metrics used to assess energy-related behaviour, the system scope at which these metrics are applied, the extent to which they are normalised, and how they are framed in terms of business-relevant concerns. To support this analysis, we define four parameters: \rqparambf{rq2.3.p1} captures the type of metric used to evaluate energy efficiency; \rqparambf{rq2.3.p2} identifies the system resource or architectural level where metrics are applied; \rqparambf{rq2.3.p3} describes how energy consumption is normalised to enable comparison across executions or configurations; and \rqparambf{rq2.3.p4} represents the business-orientation of the metric.

\begin{figure}
    \centering
    \includegraphics[width=.8\linewidth]{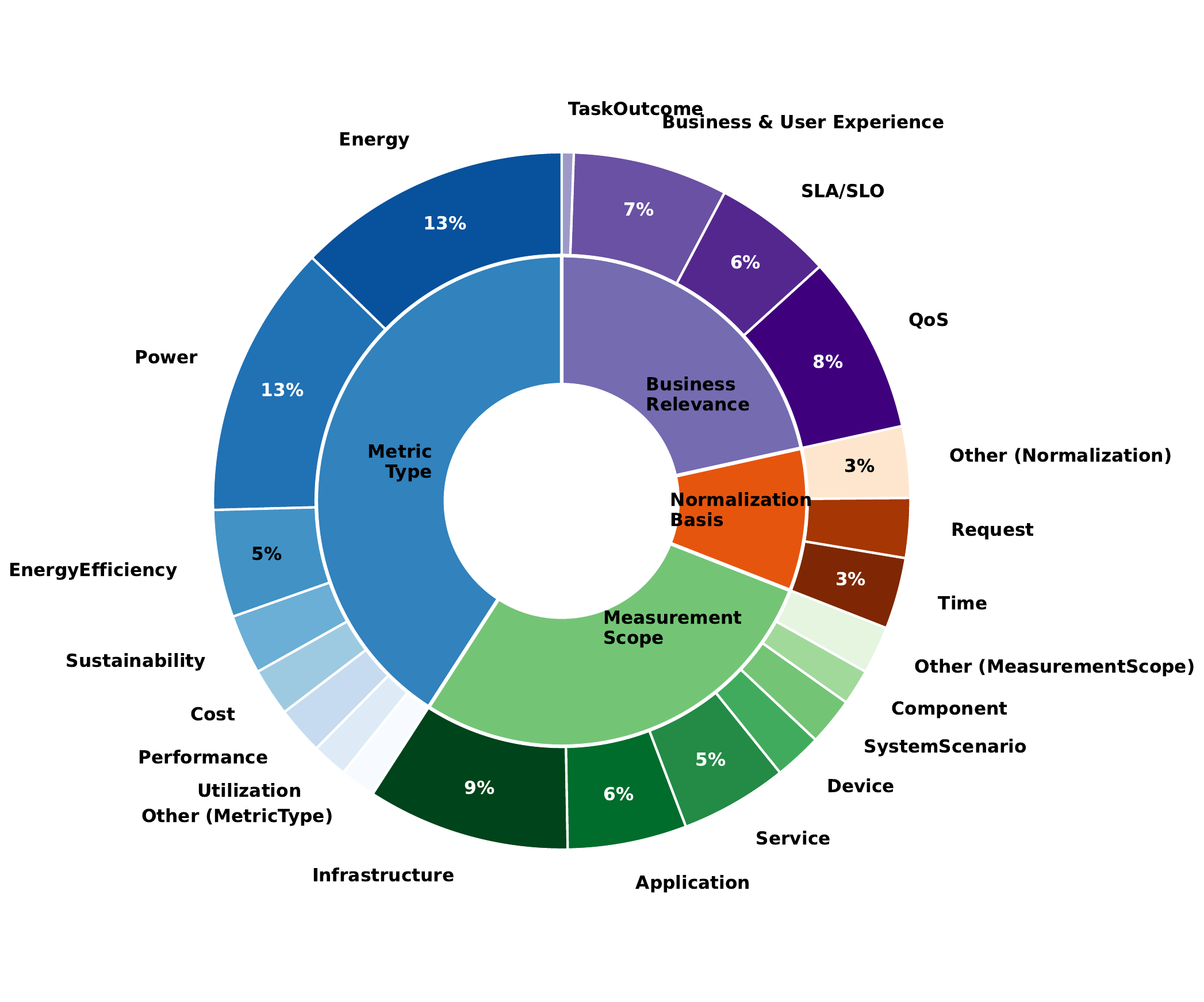}
    \caption{Combined distribution of \rqparam{rq2.3.p1}, \rqparam{rq2.3.p2}, \rqparam{rq2.3.p3}, and \rqparam{rq2.3.p4}. Percentages calculated relative to the total within each parameter category.}
    \label{fig:rq2.3_nested_combined}
\end{figure}

Figure~\ref{fig:rq2.3_nested_combined} combines the four parameters within a single visual representation. \eoan{The results show that, while energy metrics are widely reported, they are mainly defined in terms of resource consumption rather than being explicitly aligned with architectural structures or decisions, limiting their direct applicability for architecture-level reasoning.}

The distribution of \rqparam{rq2.3.p1} shows that raw energy- and power-based metrics account for over half of all reported metric types. \eoan{Efficiency-oriented metrics relate energy consumption to useful work (e.g. per request), while sustainability-oriented metrics frame energy use in environmental terms such as emissions.} This indicates that energy efficiency is mainly evaluated using direct consumption measures rather than derived or outcome-oriented indicators. In terms of \rqparam{rq2.3.p2}, energy metrics are most commonly applied at infrastructure-, application, and service-level abstractions, with fine-grained scopes such as components, containers, and functions appearing less frequently. \eoan{This suggests that, although metrics are sometimes mapped to architectural layers, they are rarely aligned with the structural units that define microservice architectures, limiting their usefulness for guiding architectural design decisions.}

\rqparam{rq2.3.p3} is reported in a relatively small subset of studies and exhibits high heterogeneity, with time- and request-based normalisation most common and several other bases appearing only sporadically. \eoan{This reflects a lack of convergence on how energy consumption should be contextualised across different system activities.} Regarding \rqparam{rq2.3.p4}, energy metrics are most often framed in relation to quality-of-service and SLA/SLO considerations, while cost- and user-oriented perspectives are present but secondary. Explicit task- or outcome-based framing appears only in isolated cases~\cite{y_huang_satedge_nodate}. 
\eoan{Overall, these findings indicate that, although energy metrics are frequently reported, they are not consistently structured in a way that directly supports architectural reasoning. Metrics tend to reflect resource usage and system behaviour rather than capturing how architectural decisions (e.g. service decomposition or communication patterns) influence energy behaviour.}

\paragraph{\textbf{Cross-parameters Analysis}}



Table~\ref{tab:rq2.3.p1_p2_mapping} maps \rqparam{rq2.3.p1} to \rqparam{rq2.3.p2}. Raw energy metrics dominate across infrastructure-, application-, and service-level observations, reflecting the widespread use of infrastructure-level monitoring tools. Power metrics are similarly concentrated at the infrastructure- and device levels, where hardware-level measurements are available. In contrast, efficiency-oriented metrics appear only in a small subset of studies and are typically associated with higher-level evaluation contexts such as component-level analysis or system-level scenarios rather than fine-grained runtime measurements. This distribution indicates that energy evaluation remains closely tied to infrastructure-level monitoring capabilities, limiting the development of metrics that directly capture architectural behaviour.

\begin{table*}[h!]
\centering
\small
\renewcommand{\arraystretch}{1.2}
\begin{tabular}{l | p{0.8cm} p{1.5cm} p{1.5cm} p{1cm} l p{1.4cm} l}
\hline
\multirow{2}{*}{\shortstack{\textbf{Measurement Scope}\\\textbf{(RQ2.3.P2)}}} &
\multicolumn{7}{c}{\textbf{Metric Type (RQ2.3.P1)}} \\
\cline{2-8}
&
\textbf{Derived Score} &
\textbf{Energy} &
\raggedright \textbf{Energy Efficiency} &
\textbf{Energy Rate} &
\textbf{Performance} &
\textbf{Power} &
\textbf{Sustainability} \\
\hline

Application &
\cite{brondolin_black-box_nodate} &
\cite{i_f_model-driven_nodate, cortellessa_v_exploring_nodate, c_courageux-sudan_studying_nodate, m_s_floroiu_anomaly_nodate, g_h_prathama_green_nodate} &
\cite{valera_energy_nodate} &
-- &
-- &
-- &
-- \\

Cluster &
-- &
\cite{z_xiang_x-man_nodate} &
-- &
\cite{m_vitali_enriching_nodate} &
-- &
-- &
-- \\

Component &
-- &
\cite{antoniou_agile_nodate} &
\cite{j_von_kistowski_teastore_nodate, dinga_empirical_nodate} &
-- &
-- &
-- &
-- \\

Device &
-- &
\cite{w_villegas-ch_adaptive_nodate, calagna_a_enabling_nodate} &
-- &
-- &
-- &
\cite{s_r_chaudhry_improved_nodate} &
-- \\

Infrastructure &
-- &
\cite{wang_l_energy-delay-aware_nodate, n_toosi_greenfog_nodate, m_xu_ibrownout_nodate, vitali_towards_nodate, h_humberto_alvarez_valera_pisco_nodate, z_bellal_gas_nodate} &
-- &
-- &
-- &
\cite{yu_y_joint_nodate, xu_m_energy_nodate-1, y_huang_satedge_nodate, agos_jawaddi_sn_analyzing_nodate, bhasi_kraken_nodate} &
\cite{n_schmitt_online_nodate} \\

Network &
-- &
\cite{khairy_simr_nodate, j_a_larracoechea_radiance_nodate} &
-- &
-- &
-- &
-- &
-- \\

Service &
-- &
\cite{v_berry_is_nodate, a_mokhtari_towards_nodate} &
\cite{k_afachao_efficient_nodate} &
-- &
-- &
-- &
-- \\

System Scenario &
-- &
\cite{saboor_a_containerized_nodate} &
\cite{c_song_service_nodate} &
-- &
\cite{e_ahvar_deca_nodate} &
-- &
-- \\

\hline
\end{tabular}
\caption{Mapping of metric types (RQ2.3.P1) across measurement scopes (RQ2.3.P2) in the reviewed studies. Only studies reporting both parameters are shown.}
\label{tab:rq2.3.p1_p2_mapping}
\end{table*}

\begin{framed} 
RQ2.3 - Energy efficiency in microservices is most commonly evaluated using raw energy and power metrics applied at infrastructure-, application-, and service-level scopes. In contrast, derived or efficiency-oriented metrics appear only in a small subset of studies and are typically associated with higher-level evaluation contexts rather than fine-grained architectural components.
\end{framed}

\subsection{Answer to the Research Question (RQ2)}
This section synthesises the findings of \textit{RQ2.1-RQ2.3} to answer \textit{"RQ2: How is energy consumption measured in microservices?"} To characterise measurement practice, we analyse three dimensions: 
measurement levels (RQ2.1), measurement tools (RQ2.2), and reported metric types (RQ2.3). Examining these dimensions together reveals where energy measurements are performed, which tools are used to obtain them, and how the resulting measurements are reported. Table~\ref{tab:rq2.1.p1_rq2.2.p1_mapping} maps \rqparam{rq2.1.p1} to \rqparam{rq2.2.p1}. The mapping highlights a strong association between measurement levels and the tools used to collect energy data. Infrastructure- and host-level measurements rely on a broad set of tools, including simulators and energy meters such as \textit{CloudSim}, \textit{SimGrid}, \textit{PISCO}, \textit{Turbostat}, \textit{RAPL}, reflecting their central role in energy measurement practice. At finer-grained levels, measurement is realised through a much narrower set of tools, including container- and kernel-level monitoring platforms (e.g., \textit{Kepler}, \textit{eBPF-based probes}) and platform-specific APIs such as \textit{Android BatteryManager}, often in combination with benchmarking or workload-generation tools.

Table~\ref{tab:rq2.1.p1_rq2.3.p1_mapping} maps \rqparam{rq2.1.p1} to \rqparam{rq2.3.p1}, showing that energy consumption is mainly reported using raw energy and power metrics across nearly all measurement levels, indicating that direct consumption measures remain the dominant reporting format. In contrast, derived metrics such as energy efficiency, energy rate, and sustainability indicators occur infrequently and are confined to a small number of infrastructure- and system-adjacent levels. Higher-level measurement does not correspond to increased use of normalised or composite metrics, suggesting that measurement granularity and metric abstraction are weakly coupled in existing studies.

\begin{table*}[t]
\centering
\small
\renewcommand{\arraystretch}{1.2}
\begin{tabular}{l | p{1.4cm} p{1.8cm} p{1.5cm} p{1cm} p{1.5cm} p{.5cm} p{1.8cm}}
\hline
\multirow{2}{*}{\shortstack{\textbf{Measurement Level} \\ \textbf{(RQ2.1.P1)}}} &
\multicolumn{7}{c}{\textbf{Tool Type (RQ2.2.P1)}} \\
\cline{2-8}
&
\textbf{Energy Estimation Tool} &
\textbf{Energy Meter} &
\textbf{Framework} &
\textbf{Model Checker} &
\textbf{Monitoring Platform} &
\textbf{OS API} &
\textbf{Simulator} \\
\hline

ClusterDatacenter &
-- &
\cite{z_xiang_x-man_nodate} &
-- &
-- &
-- &
-- &
\cite{xu_m_energy_nodate-1} \\

Component &
-- &
-- &
-- &
-- &
-- &
-- &
\cite{s_r_chaudhry_improved_nodate, n_schmitt_online_nodate} \\

ContainerPod &
-- &
-- &
-- &
-- &
\cite{a_mokhtari_towards_nodate} &
-- &
-- \\

Function &
-- &
\cite{m_s_floroiu_anomaly_nodate} &
-- &
-- &
-- &
-- &
-- \\

Hardware &
\cite{w_villegas-ch_adaptive_nodate} &
\cite{y_huang_satedge_nodate, agos_jawaddi_sn_analyzing_nodate, m_vitali_enriching_nodate, c_courageux-sudan_studying_nodate} &
-- &
-- &
-- &
-- &
\cite{cortellessa_v_exploring_nodate, calagna_a_enabling_nodate, j_a_larracoechea_radiance_nodate} \\

NodeDevice &
-- &
-- &
\cite{c_song_service_nodate} &
-- &
-- &
-- &
\cite{i_f_model-driven_nodate} \\

OS &
-- &
-- &
-- &
-- &
-- &
\cite{k_afachao_efficient_nodate} &
-- \\

Request &
-- &
\cite{v_berry_is_nodate} &
-- &
-- &
\cite{g_h_prathama_green_nodate} &
-- &
\cite{wang_l_energy-delay-aware_nodate} \\

ServerHost &
-- &
\cite{h_humberto_alvarez_valera_pisco_nodate, bhasi_kraken_nodate, valera_energy_nodate} &
-- &
\cite{z_bellal_gas_nodate} &
-- &
-- &
\cite{yu_y_joint_nodate, n_toosi_greenfog_nodate} \\

\hline
\end{tabular}
\caption{Mapping of measurement levels (RQ2.1.P1) to tool types (RQ2.2.P1). Only studies reporting both parameters are shown.}
\label{tab:rq2.1.p1_rq2.2.p1_mapping}
\end{table*}

\begin{table*}[t]
\centering
\small
\renewcommand{\arraystretch}{1.2}
\begin{tabular}{l | p{3.2cm} p{1.8cm} p{1cm} p{1.5cm} p{1cm} p{1.5cm}}
\hline
\multirow{2}{*}{\shortstack{\textbf{Measurement Level}\\\textbf{(RQ2.1.P1)}}} &
\multicolumn{6}{c}{\textbf{Metric Type (RQ2.3.P1)}} \\
\cline{2-7}
&
\textbf{Energy} &
\raggedright \textbf{Energy Efficiency} &
\textbf{Energy Rate} &
\textbf{Performance} &
\textbf{Power} &
\textbf{Sustainability} \\
\hline

Cluster Datacenter &
\cite{z_xiang_x-man_nodate} &
-- &
-- &
-- &
\cite{xu_m_energy_nodate-1} &
-- \\

Component &
-- &
-- &
-- &
-- &
\cite{s_r_chaudhry_improved_nodate} &
\cite{n_schmitt_online_nodate} \\

Container Pod &
\cite{vitali_towards_nodate, a_mokhtari_towards_nodate} &
-- &
-- &
-- &
-- &
-- \\

Function &
\cite{m_s_floroiu_anomaly_nodate} &
-- &
-- &
-- &
-- &
-- \\

Hardware &
\cite{w_villegas-ch_adaptive_nodate, saboor_a_containerized_nodate, khairy_simr_nodate, m_xu_ibrownout_nodate, cortellessa_v_exploring_nodate, calagna_a_enabling_nodate, j_a_larracoechea_radiance_nodate, c_courageux-sudan_studying_nodate} &
\cite{j_von_kistowski_teastore_nodate} &
\cite{m_vitali_enriching_nodate} &
-- &
\cite{y_huang_satedge_nodate, agos_jawaddi_sn_analyzing_nodate} &
-- \\

Node Device &
\cite{i_f_model-driven_nodate} &
\cite{c_song_service_nodate} &
-- &
-- &
-- &
-- \\

OS &
-- &
\cite{k_afachao_efficient_nodate} &
-- &
-- &
-- &
-- \\

Request &
\cite{wang_l_energy-delay-aware_nodate, v_berry_is_nodate, g_h_prathama_green_nodate} &
-- &
-- &
-- &
-- &
-- \\

Server Host &
\cite{n_toosi_greenfog_nodate, h_humberto_alvarez_valera_pisco_nodate, z_bellal_gas_nodate, antoniou_agile_nodate} &
\cite{valera_energy_nodate} &
-- &
\cite{e_ahvar_deca_nodate} &
\cite{yu_y_joint_nodate, bhasi_kraken_nodate} &
-- \\

\hline
\end{tabular}
\caption{Mapping of measurement levels (RQ2.1.P1) to metric types (RQ2.3.P1). Only studies reporting both parameters are shown.}
\label{tab:rq2.1.p1_rq2.3.p1_mapping}
\end{table*}

\begin{framed} 
RQ2 – Energy consumption in microservices is measured primarily at infrastructure- and host-adjacent levels using raw energy and power metrics. These measurements are typically obtained through a small set of widely used tools, including simulators, energy meters, and system monitoring interfaces. Fine-grained measurement at the container, request, or function level occurs less frequently, relies on more specialised tooling, and is rarely accompanied by normalised or composite efficiency metrics.
\end{framed}

\section{Architectural Solutions for Energy-Efficient Microservices (RQ3)}\label{sc:rq3}

This section addresses \textit{RQ3: What architectural solutions exist for energy-efficient microservices?} Unlike \textit{RQ1} and \textit{RQ2}, which examine where energy efficiency is considered and how it is measured, \textit{RQ3} focuses on architectural responses to energy concerns. The analysis is structured around three aspects: (i) reported architectural approaches for improving energy efficiency; (ii) trade-offs between energy efficiency and other quality attributes; and (iii) challenges affecting the adoption of energy-efficient architectural solutions.

\subsection{RQ3.1: What approaches are used to improve energy efficiency in microservice architectures?}\label{subsc:rq3.1}

This question aims to characterise how energy efficiency is addressed through architectural mechanisms, design choices, and supporting technologies within reported solutions. To support this analysis, three parameters are defined. \rqparambf{rq3.1.p1} captures the strategy through which existing systems adapt or evolve to address energy efficiency. \rqparambf{rq3.1.p2} characterises the type of patterns and tactics used to improve energy efficiency, and \rqparambf{rq3.1.p3} identifies the specific technologies that enable energy-efficient solutions.

\begin{figure}[b]
    \centering
    \includegraphics[width=.8\linewidth]{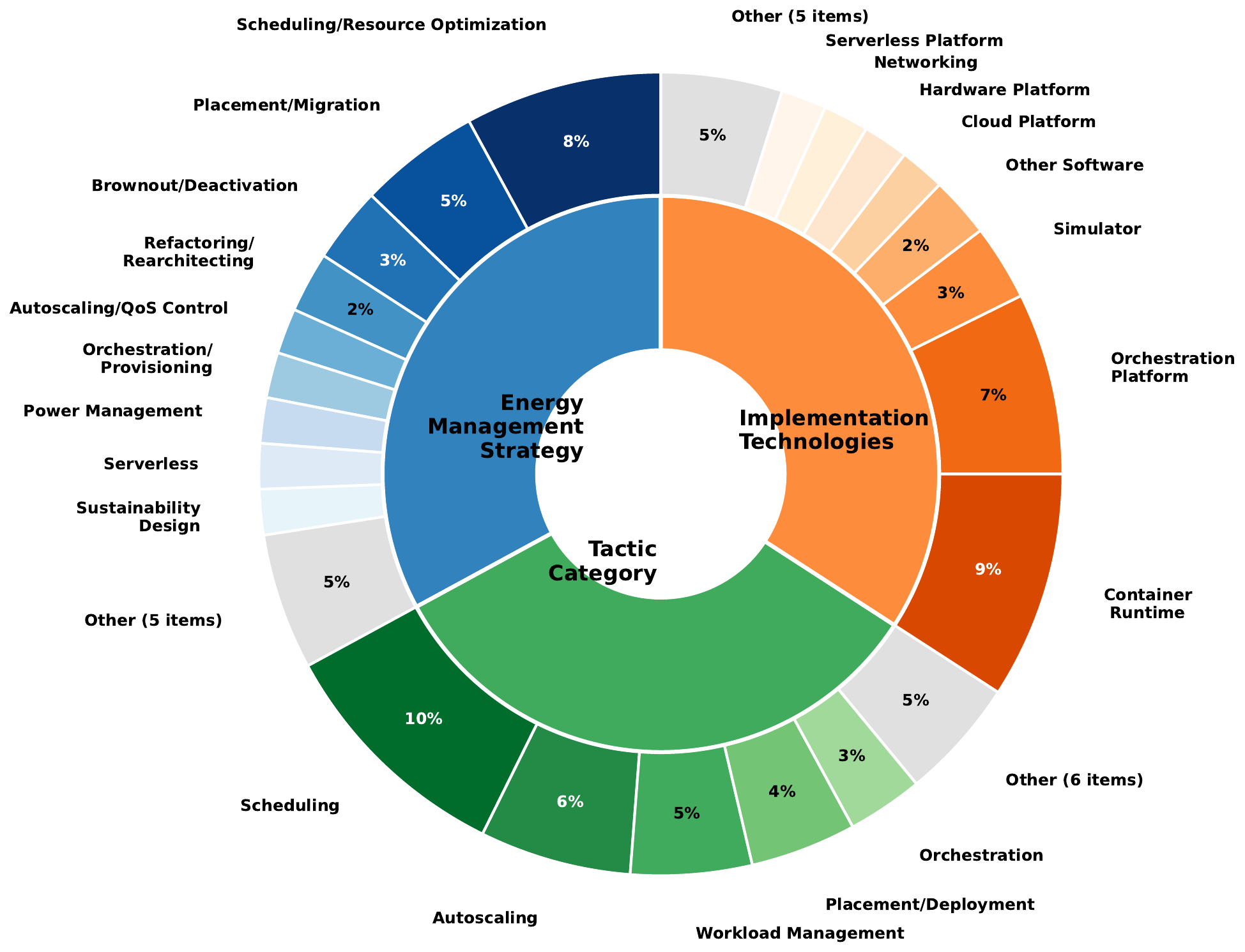}
    \caption{Combined distribution of \rqparam{rq3.1.p1}, \rqparam{rq3.1.p2} and \rqparam{rq3.1.p3} in energy-efficient microservice modernisation. Percentages calculated relative to the total within each parameter category. Values below 2\% are omitted from the visualisation for clarity.}
    \label{fig:rq3.1_nested_combined}
\end{figure}

Figure~\ref{fig:rq3.1_nested_combined} combines the three parameters within a single visual representation.
\eoan{Energy-efficient microservice modernisation is primarily driven by runtime resource optimisation strategies, realised through scheduling- and scaling-based tactics and implemented primarily on container and orchestration platforms.} \rqparam{rq3.1.p1} is strongly concentrated on resource optimisation, most commonly realised through scheduling mechanisms, with placement and migration approaches occurring to a lesser extent, while remaining strategies appear with comparatively low frequencies. \eoan{This indicates a focus on improving energy efficiency through dynamic runtime control of resource allocation, rather than through structural or design-time architectural changes.}

Across \rqparam{rq3.1.p2}, runtime-oriented tactics dominate, with scheduling-based tactics accounting for the largest share, followed by autoscaling and workload approaches. Placement, deployment, and orchestration-related tactics appear less frequently. \eoan{This reinforces the reliance on operational control mechanisms, with limited emphasis on restructuring system architecture or service composition.} The results of \rqparam{rq3.1.p3} are primarily centred on container runtimes and orchestration platforms, which account for the majority of reported cases. Simulation environments and other software-based tooling appear less frequently, while technologies such as serverless platforms, edge/IoT frameworks, or specialised optimisation tools occur only in isolated instances~\cite{gunasekaran_fifer_nodate,w_villegas-ch_adaptive_nodate,c_song_service_nodate}. \eoan{This suggests that energy optimisation is largely implemented within existing cloud-native infrastructure stacks rather than through specialised platforms.}

\begin{table*}[b]
\centering
\small
\renewcommand{\arraystretch}{1.2}
\begin{tabular}{l | p{.8cm} p{1.3cm} p{.8cm} p{1.2cm} p{2cm} p{1.2cm} p{1.2cm}}
\hline
\multirow{2}{*}{\shortstack{\textbf{Energy Management Strategy}\\\textbf{(RQ3.1.P1)}}} &
\multicolumn{7}{c}{\textbf{Tactic Category (RQ3.1.P2)}} \\
\cline{2-8}
&
\textbf{Orch.} &
\textbf{Placement / Depl.} &
\textbf{Power Mgt.} &
\textbf{Redun. / Repl.} &
\textbf{Scheduling} &
\textbf{Workload Batch.} &
\textbf{Workload Mgt.} \\
\hline

Brownout &
-- &
-- &
-- &
-- &
\cite{yu_y_joint_nodate, n_toosi_greenfog_nodate, h_humberto_alvarez_valera_pisco_nodate, bhasi_kraken_nodate, j_a_larracoechea_radiance_nodate} &
-- &
-- \\

Comm. Reconf. &
\cite{c_song_service_nodate} &
-- &
-- &
-- &
-- &
-- &
-- \\

Hardware Redesign &
-- &
-- &
-- &
-- &
-- &
\cite{cortellessa_v_exploring_nodate} &
-- \\

Offloading &
\cite{k_afachao_efficient_nodate} &
\cite{w_villegas-ch_adaptive_nodate} &
-- &
-- &
-- &
-- &
-- \\

Orch. Prov. &
-- &
-- &
-- &
-- &
\cite{wang_l_energy-delay-aware_nodate} &
-- &
-- \\

Placement Migration &
\cite{m_xu_ibrownout_nodate} &
\cite{i_f_model-driven_nodate, valera_energy_nodate} &
-- &
\cite{c_courageux-sudan_studying_nodate} &
\cite{calagna_a_enabling_nodate, n_schmitt_online_nodate} &
-- &
-- \\

Power Mgt. DVFS &
-- &
-- &
\cite{j_von_kistowski_teastore_nodate} &
-- &
\cite{v_berry_is_nodate} &
-- &
\cite{y_huang_satedge_nodate} \\

Redundancy\\Duplication &
-- &
-- &
-- &
\cite{vitali_towards_nodate} &
-- &
-- &
-- \\

Refactor &
-- &
\cite{e_ahvar_deca_nodate} &
-- &
-- &
\cite{z_xiang_x-man_nodate} &
-- &
-- \\

Sched. Opt. &
\cite{saboor_a_containerized_nodate} &
\cite{xu_m_energy_nodate-1} &
-- &
-- &
\cite{khairy_simr_nodate, m_vitali_enriching_nodate} &
-- &
\cite{g_h_prathama_green_nodate} \\

Sustain. Design &
\cite{brondolin_black-box_nodate} &
-- &
-- &
-- &
-- &
-- &
-- \\

\hline
\end{tabular}
\caption{Mapping of energy management strategy (RQ3.1.P1) to tactic categories (RQ3.1.P2). Only studies reporting both parameters are shown. Abbreviated labels are used for readability.}
\label{tab:rq3.1.p1_p2_mapping}
\end{table*}

\paragraph{\textbf{Cross-parameters Analysis}}
Table~\ref{tab:rq3.1.p1_p2_mapping} maps \rqparam{rq3.1.p1} to \rqparam{rq3.1.p2}, showing how reported modernisation strategies align with different tactic categories across the reviewed studies. The mapping indicates that scheduling-related tactic categories form the primary point of convergence, co-occurring with multiple modernisation strategies and exhibiting the broadest cross-category coverage in the dataset. Placement and deployment tactic categories also co-occur with more than one modernisation strategy, though with more limited breadth than scheduling-oriented approaches. From the perspective of modernisation strategies, placement and migration and scheduling optimisation are associated with a wider range of tactic categories, while most other strategies appear only in narrowly scoped or isolated pairings. Several tactic categories, including power management and redundancy or replication, occur only sporadically and are associated with a small number of specific modernisation strategies. These pairings do not form dominant or recurring clusters within the dataset. Overall, architectural approaches for improving energy efficiency are unevenly distributed, with a small number of runtime-focused tactic categories acting as convergence points, while most other strategy-tactic combinations occur infrequently and without systematic coupling.

\begin{framed} 
RQ3.1 - Architectural approaches for improving energy efficiency in microservices are most commonly runtime-oriented in nature, centring on scheduling- and autoscaling-oriented strategies realised through container runtimes and orchestration platforms, while architectural refactoring, power management, serverless adoption, and sustainability-by-design approaches have a limited presence in the literature.
\end{framed}

\subsection{RQ3.2: What trade-offs exist between energy efficiency and other quality attributes in microservice architectures?}\label{subsc:rq3.2}

This sub-question examines how trade-offs between energy efficiency and other quality attributes are reported. The analysis considers three dimensions: (i) the quality attributes affected; (ii) the type of relationship reported between energy efficiency and those attributes (e.g. competing, complementary, or context-dependent); and (iii) the architectural strategies used to manage or mitigate these tensions.


\begin{figure}[b]
    \centering
    \includegraphics[width=.8\linewidth]{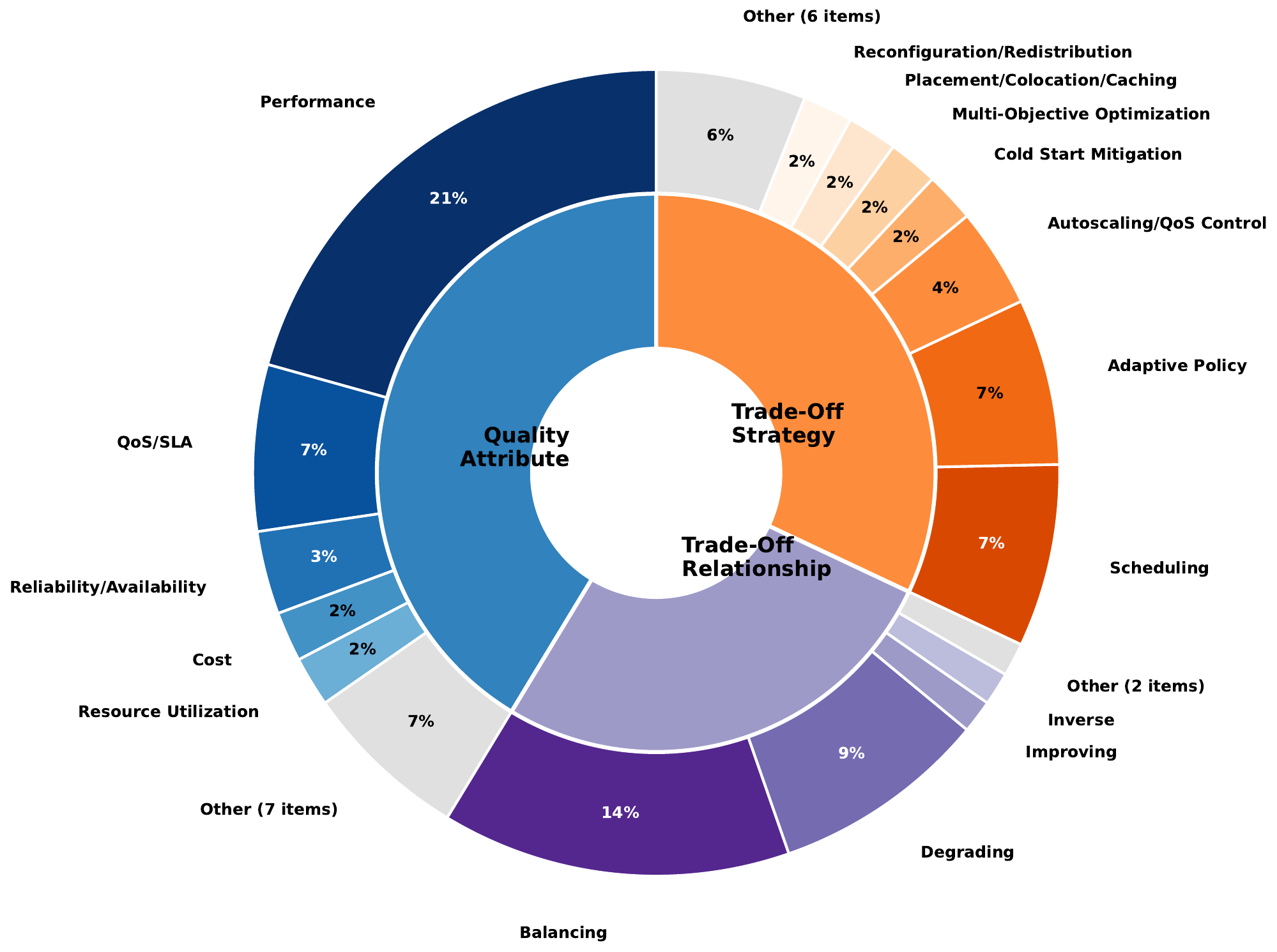}
    \caption{Combined distribution of \rqparam{rq3.2.p1}, \rqparam{rq3.2.p2}, and \rqparam{rq3.2.p3}. Percentages calculated relative to the total within each parameter category. Values below 2\% are omitted from the visualisation for clarity.}
    \label{fig:rq3.2_nested_combined}
\end{figure}

Figure~\ref{fig:rq3.2_nested_combined} combines all parameters within a single visual representation.
\eoan{Energy-related trade-offs in microservice architectures are primarily framed in terms of performance, with trade-off management largely realised through runtime control mechanisms.} Across \rqparam{rq3.2.p1}, trade-offs are centred on performance-related qualities, with quality-of-service concerns forming a secondary cluster, while other quality attributes appear far less frequently. \eoan{This indicates that energy efficiency is mainly evaluated in relation to system performance, with limited consideration of broader architectural qualities such as reliability, cost, or scalability.}

The distribution of \rqparam{rq3.2.p2} shows that trade-offs are most frequently framed as balancing relationships, followed by degrading relationships, \eoan{where energy improvements are achieved at the expense of performance (e.g. increased latency or reduced throughput).} Less common are inverse or context-dependent relationships, \eoan{where inverse denotes a consistently opposing relationship, while reversing indicates that the direction of impact changes depending on system conditions.} \eoan{This suggests that energy efficiency is typically managed as a competing objective rather than a synergistic concern.} Regarding \rqparam{rq3.2.p3}, trade-off management strategies cluster around operational control mechanisms, with scheduling- and policy-based approaches forming the main groups. \eoan{These approaches rely on runtime adaptation, reinforcing the use of operational rather than structural solutions to manage energy trade-offs.}

\begin{table*}[b]
\centering
\small
\renewcommand{\arraystretch}{1.2}
\begin{tabular}{l | l p{1.5cm} p{1.2cm} l p{1cm} l}
\hline
\multirow{2}{*}{\shortstack{\textbf{Trade-off Relationship}\\\textbf{(RQ3.2.P2)}}} &
\multicolumn{6}{c}{\textbf{Quality Attribute (RQ3.2.P1)}} \\
\cline{2-7}
&
\textbf{Cost} &
\raggedright \textbf{Energy Efficiency} &
\textbf{Output Quality} &
\textbf{Performance} &
\textbf{QoS/SLA} &
\textbf{User Experience} \\
\hline

Degrading &
\cite{e_ahvar_deca_nodate} &
-- &
\cite{g_h_prathama_green_nodate} &
\cite{a_mokhtari_towards_nodate, brondolin_black-box_nodate, m_s_floroiu_anomaly_nodate, antoniou_agile_nodate} &
\cite{z_xiang_x-man_nodate} &
\cite{yu_y_joint_nodate} \\

ExplicitTradeOff &
-- &
\cite{k_afachao_efficient_nodate} &
-- &
-- &
-- &
-- \\

Improving &
-- &
-- &
-- &
\cite{w_villegas-ch_adaptive_nodate, khairy_simr_nodate} &
-- &
-- \\

Inverse &
-- &
\cite{c_song_service_nodate} &
-- &
\cite{calagna_a_enabling_nodate} &
-- &
-- \\

Reversing &
-- &
-- &
-- &
\cite{c_courageux-sudan_studying_nodate} &
-- &
-- \\

\hline
\end{tabular}
\caption{Mapping of trade-off relationships (RQ3.2.P2) to impacted quality attributes (RQ3.2.P1). Only studies reporting both parameters are shown.}
\label{tab:rq3.2.p2_p1_mapping}
\end{table*}

\paragraph{\textbf{Cross-parameters Analysis}}
Table~\ref{tab:rq3.2.p2_p1_mapping} maps \rqparam{rq3.2.p2} to \rqparam{rq3.2.p1}, showing how different trade-off relationships are associated with reported quality attributes. The mapping indicates that performance acts as the primary point of convergence across reported trade-off relationships, appearing across multiple relationship types. Degrading trade-offs exhibit the broadest associations, co-occurring with a wider range of quality attributes. In contrast, other trade-off relationships show more constrained patterns, with associations limited to a small subset of quality attributes and, in several cases, to single attribute categories. From the perspective of quality attributes, non-performance-related qualities are associated with a narrower set of trade-off relationships and do not recur across multiple relationship types. Overall, reported trade-offs between energy efficiency and other qualities are unevenly distributed, with broad associations concentrated in a small number of relationship types, while most relationships are characterised by limited and tightly scoped pairings.

\begin{framed} 
RQ3.2 - Reported trade-offs between energy efficiency and other quality attributes in microservices are centred on performance-related concerns and are most often characterised as balancing or degrading relationships, with trade-offs typically described as moderate in severity and managed through operational control strategies, while other quality attributes and trade-off forms appear in more limited and narrowly scoped cases.
\end{framed}


\subsection{RQ3.3: What challenges affect the adoption of energy-efficient architectural solutions in microservices?}\label{subsc:rq3.3}
This question focuses on identifying the types of challenges encountered, the lifecycle phases at which they arise, and how their complexity is characterised in reported studies. To support this analysis, three parameters are defined. \rqparambf{rq3.3.p1} captures the type of challenge reported in relation to adopting energy-efficient architectural solutions. \rqparambf{rq3.3.p2} identifies the lifecycle phase at which the reported challenge occurs, and \rqparambf{rq3.3.p3} characterises the reported complexity.

\begin{figure}[t!]
    \centering
    \includegraphics[width=.8\linewidth]{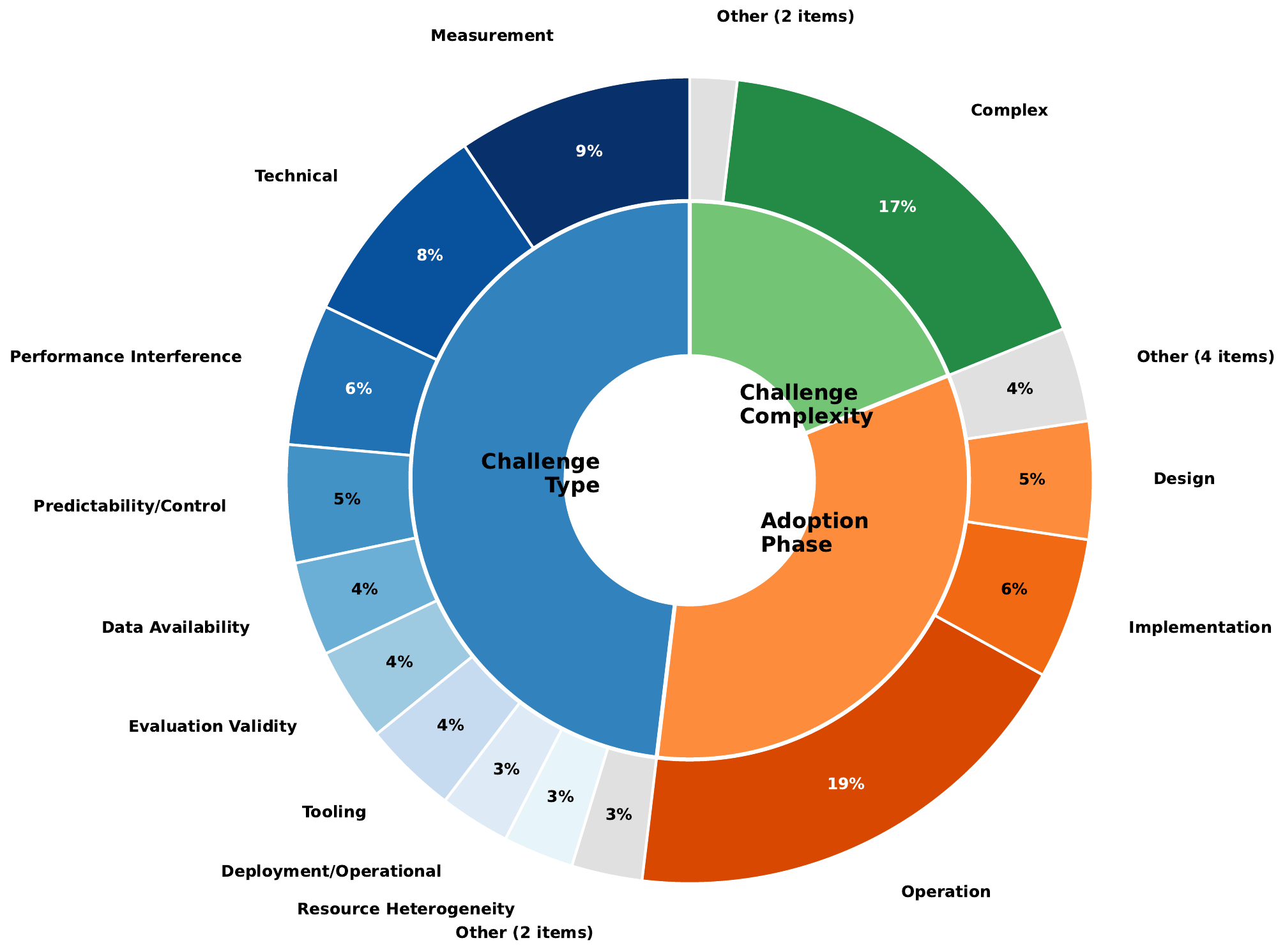}
    \caption{Combined distribution of \rqparam{rq3.3.p1}, \rqparam{rq3.3.p2}, and \rqparam{rq3.3.p3}. Percentages calculated relative to the total within each parameter category. Values below 2\% are omitted from the visualisation for clarity.}
    \label{fig:rq3.3_nested_combined}
\end{figure}

Figure~\ref{fig:rq3.3_nested_combined} combines all parameters within a single visual representation.
\eoan{Reported challenges in energy-efficient microservice modernisation are primarily technical and measurement-related, occur predominantly during system operation, and are consistently characterised as complex.} Across \rqparam{rq3.3.p1}, reported challenges are centred on measurement and technical concerns, followed by performance interference and predictability or control, while other challenge types appear less frequently and do not form recurring patterns. \eoan{This indicates that the primary barriers to energy efficiency lie in accurately measuring and managing energy behaviour within distributed systems, rather than in higher-level organisational or design concerns.}

The distribution of \rqparam{rq3.3.p2} shows that challenges are concentrated during system operation, with fewer reported during implementation and design phases, and other lifecycle phases appearing only in isolated cases \eoan{\cite{valera_energy_nodate,i_f_model-driven_nodate,z_bellal_gas_nodate}}. \eoan{This suggests that energy-related challenges emerge mainly during runtime, where system behaviour, workload variability, and resource interactions become visible.} For \rqparam{rq3.3.p3}, reported challenges are overwhelmingly characterised as complex, with only isolated deviations. \eoan{This reflects the inherent difficulty of managing energy efficiency in microservice systems, where interactions across distributed components and dynamic environments introduce significant uncertainty.}


\begin{table*}[t]
\centering
\small
\renewcommand{\arraystretch}{1.2}
\begin{tabular}{l | llllll}
\hline
\multirow{2}{*}{\textbf{Challenge Type (RQ3.3.P1)}} &
\multicolumn{6}{c}{\textbf{Adoption Phase (RQ3.3.P2)}} \\
\cline{2-7}
&
\textbf{Design} &
\textbf{Evaluation} &
\textbf{Implementation} &
\textbf{Operation} &
\textbf{Planning} &
\textbf{Production} \\
\hline

Deployment / Operational &
-- &
-- &
-- &
\cite{j_von_kistowski_teastore_nodate, valera_energy_nodate} &
-- &
-- \\

Evaluation Validity &
\cite{g_h_prathama_green_nodate} &
-- &
-- &
\cite{c_song_service_nodate} &
\cite{vitali_towards_nodate} &
\cite{j_a_larracoechea_radiance_nodate} \\

Measurement &
\cite{brondolin_black-box_nodate} &
\cite{v_berry_is_nodate} &
\cite{w_villegas-ch_adaptive_nodate} &
\cite{agos_jawaddi_sn_analyzing_nodate, m_s_floroiu_anomaly_nodate, antoniou_agile_nodate} &
-- &
-- \\

Organizational &
\cite{dinga_empirical_nodate} &
-- &
-- &
-- &
-- &
-- \\

Predictability / Control &
-- &
-- &
\cite{y_huang_satedge_nodate} &
\cite{khairy_simr_nodate, h_humberto_alvarez_valera_pisco_nodate} &
-- &
-- \\

Programmability &
-- &
-- &
\cite{cortellessa_v_exploring_nodate} &
-- &
-- &
-- \\

Technical &
\cite{xu_m_energy_nodate-1, a_mokhtari_towards_nodate} &
-- &
-- &
\cite{z_xiang_x-man_nodate, wang_l_energy-delay-aware_nodate, i_f_model-driven_nodate, bhasi_kraken_nodate} &
-- &
-- \\

\hline
\end{tabular}
\caption{Mapping of reported challenge types (RQ3.3.P1) to adoption phases (RQ3.3.P2). Only studies reporting both parameters are shown.}
\label{tab:rq3.3.p1_p2_mapping}
\end{table*}

\paragraph{\textbf{Cross-parameters Analysis}}
Table~\ref{tab:rq3.3.p1_p2_mapping} maps \rqparam{rq3.3.p1} to \rqparam{rq3.3.p2}. Although many challenges are reported during operation, some emerge earlier during design and implementation, indicating that barriers to adopting energy-efficient microservices can arise before deployment. However, these early-phase associations are isolated and do not form consistent patterns. Overall, challenges span the lifecycle, but are primarily examined at runtime.

\begin{framed} 
RQ3.3 - Reported challenges affecting the adoption of energy-efficient architectural solutions in microservices are primarily operational in nature and are most often characterised as complex, with a smaller number of challenges also reported during design and implementation phases, while other adoption phases and challenge types occur only in limited and isolated cases.
\end{framed}

\subsection{Answer to the Research Question (RQ3)}
This section synthesises the findings of \textit{RQ3.1-RQ3.3} to answer \textit{“RQ3: What architectural solutions exist for energy-efficient microservices?”}. Across reported modernisation strategies, architectural solutions for improving energy efficiency are realised primarily through operational and runtime-oriented mechanisms rather than through pervasive architectural restructuring. Reported solutions focus on controlling system behaviour during execution, with comparatively limited emphasis on architectural redesign or lifecycle-spanning integration of energy-efficiency concerns. The reported use of these solutions is accompanied by trade-offs with other quality attributes, which are most often framed as the need to balance competing system objectives. Such trade-offs are generally moderate and are managed through operational strategies aligned with the runtime nature of the solutions. Adoption challenges further reinforce this pattern. Reported challenges are largely encountered during system operation and are commonly described as complex, with fewer challenges reported at earlier lifecycle stages and little differentiation beyond the dominant operational focus.

\begin{framed} 
RQ3 - Architectural solutions for energy-efficient microservices reported in the literature are most commonly runtime-oriented and operational in nature, with energy efficiency addressed through execution-time control mechanisms rather than through structural architectural redesign. These solutions are typically associated with moderate trade-offs and complex operational-phase adoption challenges, while broader architectural integration of energy efficiency remains limited.
\end{framed}
\section{Discussion}\label{sc:discussion}


This section synthesises the findings of \textit{RQ1}-\textit{RQ3} to identify broader structural patterns in research on energy-efficient microservices. The results collectively show how energy is positioned within microservice architectures in terms of lifecycle integration, measurement practices, and architectural solutions.

\subsection{Runtime-Centrism in Energy-Efficient Microservices}\label{subsc:runtime-centrism}

The synthesis of \textit{RQ1}-\textit{RQ3} indicates that energy efficiency in microservices is largely framed as a runtime optimisation concern. Across lifecycle analysis, measurement practices, architectural solutions, and trade-offs, energy is typically introduced after deployment and addressed through execution-time mechanisms such as scheduling, autoscaling, and resource reconfiguration. Architectural reasoning, therefore, focuses on optimising energy consumption within existing system structures rather than shaping those structures through energy-aware design. While this runtime-oriented approach aligns with the dynamic nature of microservice environments, it limits architectural influence on energy efficiency. Structural decisions such as service decomposition, interaction topology, and data placement are rarely evaluated from an energy perspective and are usually fixed before energy considerations are introduced. Advancing the field requires integrating energy reasoning earlier in the lifecycle, including during microservice decomposition, service boundary design, and deployment topology evaluation. Treating energy efficiency as a structural design dimension rather than solely a runtime optimisation parameter represents an important direction for future research.

\begin{framed} 
Takeaway 1 - Current research treats energy efficiency primarily as a runtime optimisation problem layered on top of existing microservice architectures. Extending energy considerations into early architectural decision-making remains a central research opportunity.
\end{framed}

\subsection{Measurement Granularity and Architectural Solution Space}\label{subsc:measurement-granularity}

While runtime-centrism characterises where energy efficiency is addressed, the findings of \textit{RQ2} and \textit{RQ3} indicate that measurement granularity strongly shapes the architectural solution space. Energy consumption is most often measured at infrastructure-adjacent levels, such as servers, nodes, or container runtimes, using monitoring tools that provide coarse-grained visibility. Fine-grained measurement at the request, function, or interaction level is less common and typically relies on specialised tools. This measurement profile aligns with the dominant solution categories identified in \textit{RQ3.1}. Approaches such as scheduling, autoscaling, placement, and orchestration operate at infrastructure or deployment levels that correspond to the resolution of available measurements. Strategies requiring detailed service-internal or interaction-level energy attribution are comparatively rare. This pattern suggests that methodological constraints shape architectural exploration. When energy signals are available primarily at node or container levels, optimisation tends to focus on those levels. Advancing architectural energy reasoning requires closer integration between measurement mechanisms and architectural abstractions, including attribution models that link low-level energy signals to architectural constructs and tracing frameworks that associate energy usage with service interactions.

\begin{framed} 
Takeaway 2 - The granularity of available energy measurement constrains the architectural solution space explored in the literature. Expanding fine-grained attribution mechanisms represents a key enabler for structurally grounded energy-aware design.
\end{framed}

\subsection{Performance-Centric Trade-offs and Architectural Implications}\label{subsc:performance-centric}

Across the reviewed literature, energy efficiency is rarely positioned as a primary architectural objective. Instead, it is typically discussed in relation to performance and service-level trade-offs, where it is balanced against latency, throughput, or quality-of-service constraints. Within this framing, energy optimisation does not redefine architectural priorities but is negotiated within an existing performance-oriented optimisation paradigm. As a result, architectural decisions rarely reconsider service boundaries, decomposition strategies, or interaction models. Energy concerns are instead addressed through adjustments that preserve established performance objectives, such as runtime tuning, resource allocation, or scheduling mechanisms. This performance-centric framing narrows the scope of energy-aware architectural reasoning. When energy is evaluated mainly through execution-time metrics, architectural assessment tends to prioritise responsiveness over longer-term sustainability considerations. Advancing the field requires treating energy efficiency as a first-class system quality attribute. Integrating energy into multi-attribute evaluation frameworks and architectural decision-making methods would enable sustainability considerations to influence structural design choices rather than remaining a secondary optimisation concern.

\begin{framed} 
Takeaway 3 - Energy efficiency is predominantly framed as a performance trade-off rather than as a primary architectural objective. Elevating energy to a first-class quality attribute represents a key step toward structurally grounded sustainable microservice design.
\end{framed}

\subsection{Toward Maturity of Metrics and Measurement Methods}\label{subsc:metric-maturity}




The findings of \textit{RQ2.3} indicate that energy evaluation is largely based on simple measurement practices relying on raw energy and power measurements reported at infrastructure or container-adjacent levels. Few studies employ workload-oriented metrics such as energy per request or energy per unit of functionality, which relate energy consumption to system behaviour. This metric profile limits comparability across studies and weakens the empirical grounding for architectural reasoning. When energy measurements are reported mainly as absolute values tied to specific infrastructures or workloads, it becomes difficult to evaluate how architectural decisions influence energy efficiency across different contexts. Advancing the field requires greater methodological maturity in energy evaluation. Promising directions include standardised benchmarking scenarios for energy-aware microservices, broader adoption of workload-oriented metrics, and stronger integration of sustainability indicators into architectural evaluation frameworks.

\begin{framed} 
Takeaway 4 - Energy evaluation in microservice research is dominated by raw infrastructure-level metrics, limiting cross-study comparability and architectural interpretation. Advancing methodological maturity is essential for enabling robust, architecture-informed sustainability reasoning.
\end{framed}

\subsection{Implications for Practitioners}\label{subsc:practitioners}
The findings of this review have direct implications for software architects and engineers designing microservices:

\begin{enumerate}
    \item Energy efficiency should be considered during architectural design, \eoan{as current research mainly addresses it through runtime optimisation mechanisms such as monitoring, scheduling, and autoscaling~\cite{gunasekaran_fifer_nodate, n_toosi_greenfog_nodate}.} Decisions regarding service decomposition, interaction topology, and deployment structure influence long-term energy behaviour and cannot be fully addressed solely through autoscaling or scheduling.
    \item Measurement capabilities should align with architectural abstractions. Reliance on node- or container-level metrics~\cite{z_xiang_x-man_nodate,calagna_a_enabling_nodate} limits visibility into how services and interactions contribute to overall consumption. Practitioners should prioritise observability infrastructure that supports fine-grained attribution of energy usage.
    \item Energy efficiency should be treated as a first-class quality attribute in architectural design and evaluation, \eoan{particularly in light of its interaction with performance and other system qualities observed in the reviewed studies.} Framing energy only as a performance trade-off risks marginalising sustainability considerations. Explicitly incorporating energy into architectural decision-making can support more balanced trade-off analysis.
    \item Metric design requires careful consideration. Reporting raw energy or power values alone limits comparability and practical interpretation. Workload-oriented metrics, such as energy per request or per unit of functionality~\cite{a_mokhtari_towards_nodate,h_humberto_alvarez_valera_pisco_nodate}, support more meaningful evaluation across workloads and deployment contexts.
\end{enumerate}
\section{Threats to Validity} \label{sc:threats}
This study is subject to threats to validity inherent to systematic literature reviews. The threats are discussed following the classification for secondary studies proposed by Ampatzoglou et al.~\cite{ampatzoglou2019identifying}.

\textbf{Study selection validity} concerns threats related to identifying and including relevant primary studies. The search string required explicit references to microservices and energy-related terms, ensuring focus on their intersection, but may have excluded studies addressing related concerns (e.g., sustainability, resource optimisation) using different terminology. Backwards and forward snowballing were applied to identify additional studies; however, the same criteria were used to maintain consistency, so studies that did not explicitly refer to microservices or energy may still have been excluded. The screening process combined manual and automated rounds, including LLM-assisted primary screening. While improving scalability and consistency, automated filtering may introduce misclassification risks depending on prompt design and model behaviour. Secondary and final screening were conducted by one author, with consultation in uncertain cases, which may introduce selection bias due to the absence of fully independent double-coding.

\textbf{Data validity} concerns the correctness and consistency of the extracted dataset and interpretation of the analysed studies. The parameter schema (e.g. lifecycle stage awareness, energy integration point) was derived from prior research experience and refined iteratively during screening and data extraction. The schema reflects the authors’ conceptual framing of architectural energy efficiency, and alternative parameterisations (e.g. carbon intensity, embodied energy) could lead to different classifications. To mitigate this risk, parameter definitions were documented, applied consistently, and discussed among the authors to resolve ambiguities.

\textbf{Research validity} concerns threats related to the overall design and scope of the review. The review focuses on peer-reviewed literature published since 2015. In addition, many included studies evaluate energy efficiency in cloud-native, containerised, or edge environments. Findings may therefore reflect research contexts rather than industrial practice, and characterise the state of published research rather than the full spectrum of industry adoption.

This study followed established SLR guidelines and defined explicit research questions, screening criteria, parameter schemas, and scoring procedures. A publicly available replication package provides instructions, search strings, screening prompts, data extraction templates, and scoring procedures~\cite{odea2026slrreplication}. While minor variations may arise due to changes in digital library indexing or LLM behaviour over time, the availability of these artefacts supports independent replication.
\section{Conclusion}\label{sc:conclusion}

This systematic literature review examines how energy efficiency is addressed in microservices from an architectural perspective. Across 40 primary studies, the evidence shows a clear imbalance: energy efficiency is widely addressed through runtime monitoring, scheduling, and autoscaling, yet rarely embedded as a first-class architectural concern. Most studies optimise energy within existing system structures rather than shaping those structures around energy-aware principles. Measurement practices remain largely infrastructure-centric and estimation-driven, and trade-offs with performance are managed reactively rather than through systematic architectural reasoning. These patterns suggest that research has progressed in tooling and runtime optimisation, but has not yet matured into a cohesive architectural discipline. Energy efficiency is often treated as an adjunct rather than an integral part of architectural design. By consolidating lifecycle positioning, measurement practices, solution strategies, and trade-off reasoning, this review clarifies how existing research fragments energy concerns across operational, infrastructural, and architectural dimensions. Advancing the field requires earlier design-stage integration, explicit modelling of cross-service energy effects, and decision-support approaches that treat energy alongside performance, scalability, and reliability in architectural trade-off analysis.

\section*{Acknowledgements}
The authors used generative AI tools to assist with writing this paper: spelling, grammar, and readability improvements. All technical ideas, analyses, results, and conclusions in this paper were conceived, developed, and verified solely by the authors. The authors take full responsibility for the content of the final manuscript.



\bibliographystyle{ACM-Reference-Format}
\bibliography{final-dataset}


\end{document}